\documentclass[epj]{svjour}

\usepackage{latexsym}

\usepackage{subfigure}
\usepackage{graphicx}
\usepackage{amssymb}
\usepackage{amsmath}
\usepackage{amsbsy}
\usepackage{color}
\graphicspath{{figures/}}
\usepackage[numbers, square, comma, sort&compress]{natbib}
\renewcommand{\vec}[1]{\boldsymbol{#1}}
\begin{document}\sloppy 
\title{Quantitative Comparison Between Crowd Models for Evacuation Planning and Evaluation}
\titlerunning{Quantitative Comparison Between Crowd Models}
\author{Vaisagh Viswanathan\inst{1}
\and Chong Eu Lee\inst{2}
\and Michael Harold Lees\inst{1, 3, 4}
\and Siew Ann Cheong\inst{2, 4}
\and Peter M. A. Sloot\inst{1, 3, 4, 5}
}                     
%
%

\institute{School of Computer Engineering, Nanyang Technological University, Nanyang Avenue, Singapore 639798, Republic of Singapore
\and
Division of Physics and Applied Physics, School of Physical and Mathematical Sciences, Nanyang Technological University, 21 Nanyang Link, Singapore 637371, Republic of Singapore
\and
Computational Science, University of Amsterdam, Science Park 904, Amsterdam, the Netherlands
\and
Complexity Program, Nanyang Technological University, 60 Nanyang View, Singapore 639673, Republic of Singapore
\and
National Research Institute ITMO, St. Petersburg, Russia
}
\date{Received: date / Revised version: date}
%
\abstract{
Crowd simulation is rapidly becoming a standard tool for evacuation planning and evaluation. However, the many crowd models in the literature are structurally different, and few have been rigorously calibrated against real-world egress data, especially in emergency situations. In this paper we describe a procedure to quantitatively compare different crowd models or between models and real-world data. We simulated three models: (1) the lattice gas model, (2) the social force model, and (3) the RVO2 model, and obtained the distributions of six observables: (1) evacuation time, (2) zoned evacuation time, (3) passage density, (4) total distance traveled, (5) inconvenience, and (6) flow rate. We then used the DISTATIS procedure to compute the compromise matrix of statistical distances between the three models. Projecting the three models onto the first two principal components of the compromise matrix, we find the lattice gas and RVO2 models are similar in terms of the evacuation time, passage density, and flow rates, whereas the social force and RVO2 models are similar in terms of the total distance traveled. Most importantly, we find that the zoned evacuation times of the three models to be very different from each other. Thus we propose to use this variable, if it can be measured, as the key test between different models, and also between models and the real world. Finally, we compared the model flow rates against the flow rate of an emergency evacuation during the May 2008 Sichuan earthquake, and found the social force model agrees best with this real data.
} 
\maketitle
\section{Introduction}
\label{intro}

Crowd simulation is an area that has been the subject of a significant amount of multidisciplinary work over the last few decades~\cite{Still:2000tp,Zhou:2010:CMS:1842722.1842725,Gwynne1999741}. Its applications range from simulating crowds for movies~\cite{Regelous:2011vt,Reynolds:1987vm} and games~\cite{Snape:2012,ageOfEmpires:2013} to analyzing pedestrian behavior~\cite{PhysRevE.51.4282,Viswanathan:ut,Guy:2010uv} and preparing for fire evacuations and similar emergencies~\cite{Klupfel:2005to,PEDFull:2011,Mordvintsev:2012}. The earliest attempts to simulate crowds generally adopted a macroscopic approach~\cite{Henderson:1974ve,WattsJr:1987tx}, where there is no explicit notion of an individual.
Later, with increasing computational resources and with availability of observational data on an individual level~\cite{CGF:CGF1090, HuNan2013, JOHANSSON2008}, modelers were able to develop microscopic approaches~\cite{Reynolds:1987vm,PhysRevE.51.4282} for application in areas where it was necessary to model and analyze individuals in the crowd.
For example, in a simulation of evacuation, knowledge of the movement of crowds could reveal methods to improve crowd flow and evacuation speed.

One of the earliest and seminal works in individual-based motion planning was Craig Reynolds's model of coordinated animal motion such as bird flocks and fish schools~\cite{Reynolds:1987vm}. Okazaki and Matsushita~\cite{Okazaki:1993wh} assigned magnetic poles to goals, agents and obstacles to model movement. Subsequently, Helbing's Social Force Model~\cite{PhysRevE.51.4282} was developed, which is still one of the most popular models of movement in crowds. More recently, there have been several velocity-based approaches to motion planning, such as the synthetic vision based model~\cite{Ondrej:2010hv} and the Reciprocal Velocity Obstacle Model~\cite{vandenBerg:2011ww,Guy:2010ko,vandenBerg:2008fu}.


The advocation of simulation-based analysis has become increasingly common over the last decade. Some well know applications include analysis of the yearly Muslim Hajj~\cite{Hughes:2003}, or more recently the Love Parade disaster, Germany 2010~\cite{Helbing:2012}. In the case of the latter, models and expertise had been used to guarantee the safety of the event, only for unforeseen circumstances to result in the deaths of 21 individuals. Clearly these real world examples emphasize the critical role that crowd modelling plays in safety preparation and planning, this in turn emphasizes the need for understanding the model dynamics, limitations and similarities. The extent to which these models are capable of accurately predicting the motion of a crowd is therefore critical for planning and safety.

Ideally these models should be validated against real-world data from all scenarios, for all cultures and for all varieties of crowd composition. Unfortunately real world data regarding egress or emergency situations is limited, often incomplete (the initial conditions are hard to know) and certainly not controlled. Even in the rare circumstances where data is available this often describes measurements and phenomena at the macro-scale, e.g., flow, density, average speed, etc. Because these models of individuals exhibit emergent behaviors at the mesoscopic and macroscopic scales, it is in general hard to tell whether the dynamics at the individual level are correct. The best researchers can realistically hope for is to collect microscopic data for a single scenario to calibrate the model. An alternative approach to the data-intensive one is to instead look quantitatively at the fundamental dynamics of the models and understand where these models differ, at the same time identifying common aspects of the models. This will aid in understanding fundamental aspects of all crowd models and offer insight into real-world dynamics of human crowds.

While much of the existing research does offer basic comparison of proposed models with existing models to demonstrate their usefulness, there is no existing quantitative comparison of the differences and similarities of these models to tell which is most accurate. The objective of this paper is to demonstrate a methodology for quantitative comparison of simulation models in a simple egress simulation. For this, we choose three popular models that are structurally very different. Since our focus is on the comparison methodology, we do not worry about specific versions of the models, in particular recent modifications and enhancements that are supposed to produce more realistic crowd behaviors.

The contribution of the paper is then three-fold: firstly, the analysis and comparison of these models provides interesting insight into the consequence of adopting each in particular forms of crowd or egress simulation. Secondly, the systematic approach we describe could be used in future to compare further models and develop a standard method of comparison for crowd simulation. Finally, the paper conclusion identifies a single measurable metric that is most effective in distinguishing the behaviour of the models.

The remainder of this paper is organised as follows. The models that are consider in the comparison are first described in Section~\ref{Models}; Section~\ref{Methodology} describes the experiments that were conducted and the methodology for analysis. Significant observations from the simulations are presented and analyzed in Section~\ref{Discussion}. Finally, Section~\ref{Conclusions} concludes the paper.

\section{Models}
\label{Models}

In this section we describe the three individual (or microscopic) models that are compared in this paper. We use the general term agent to refer to individuals within each of the models. The lattice gas model is a probabilistic approach where the future location of an agent is probabilistically determined based on the current configuration of its neighborhood. This implies that the same initial configuration can produce different results during different runs. On the other hand, in the social force model and reciprocal velocity obstacle model, agents find their ways to their destinations via deterministic collision-avoiding calculations. As a result these models produce the same result on multiple runs for a particular initial configuration.

\subsection{Lattice gas model}
\label{LatticeGasModel}

A Cellular Automation (CA) model is one in which space and time are discrete. Furthermore, the state space of a CA model is also discrete and finite. In each time step the values of all cells are updated synchronously based on the values of cells in their neighborhoods. Depending on the type of neighborhood (i.e., von Neumann, Moore), and the type of lattice (triangular, square, hexagonal, etc.), the exact number of cells in the neighborhood of a given cell can vary~\cite{Hoekstra:2010}.
Lattice gas models are CA models that make use of a discretized version of the Boltzmann transport equation to model motion~\cite{Marconi:2002ue,Marconi2002,Nagai:2004kl}. Advances in this modeling area include the Extended Floor Field Model~\cite{nishinari2004extended} in which agents interact through virtual traces that act like the pheromones in chemotaxis, and the SWARM information model~\cite{Henein:2006jq} which uses multiple floor fields to model transmission of knowledge between agents.
In this study we use the model by Tajima and Nagatani~\cite{Tajima:2001to}. In contrast to later CA models with additional features like bi-directional movement~\cite{isobe2004experiment} and vision impairment during evacuation~\cite{Nagai:2004kl}, this original model is simple, and adequate for our model comparison purpose.

\begin{figure*}[!htb]
\centering
\includegraphics[width=.8\textwidth]{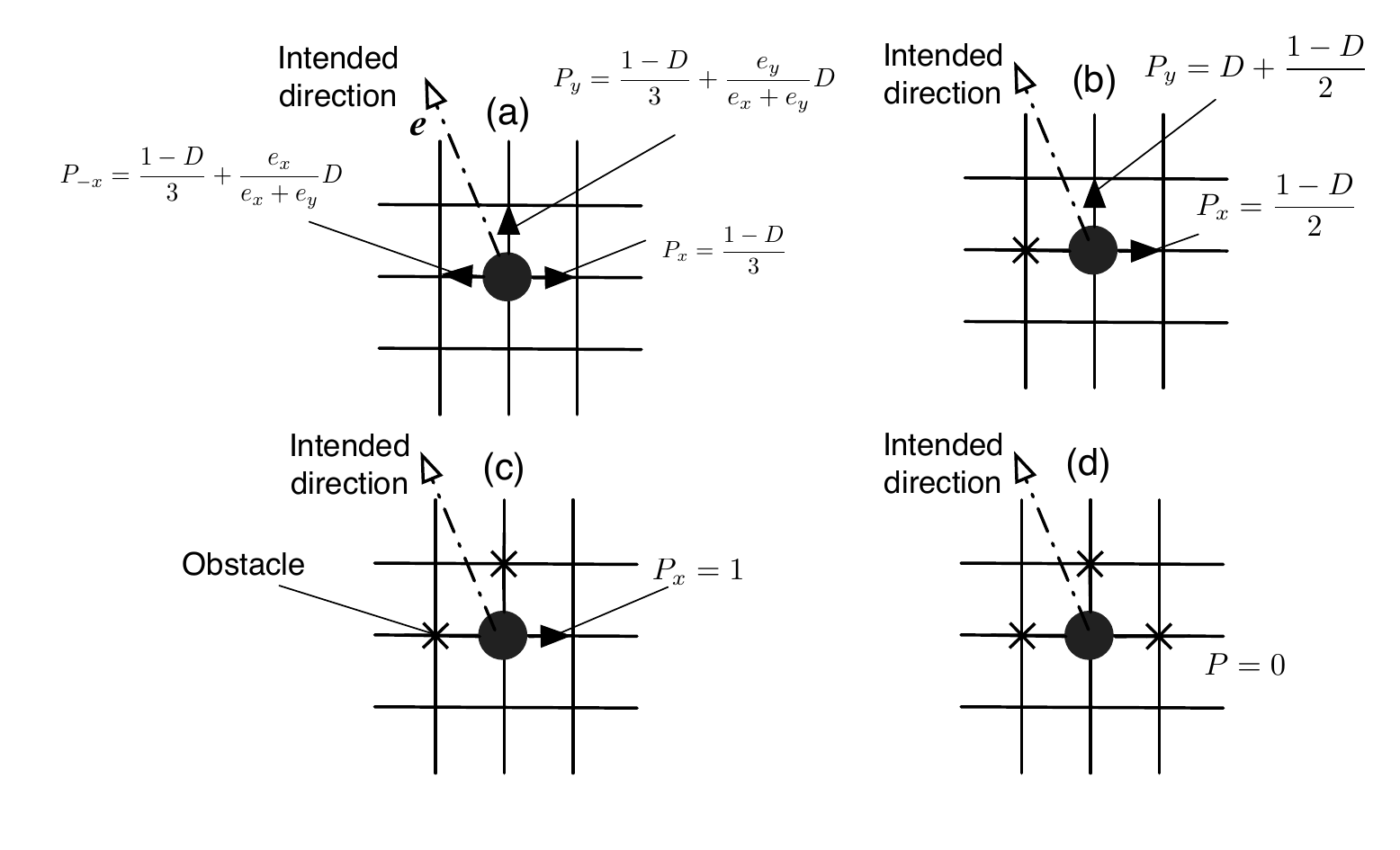}
\caption{Four out of eight possible configurations of a walker on the square lattice moving towards an exit in the direction $\vec{e}$ shown: (a) unobstructed walker, (b) walker obstructed to the left, (c) walker obstructed to the left and top, and (d) completely obstructed walker.}
\label{fig:latticeGasMovement}
\end{figure*}

A square grid is used, and each cell
can be occupied by at most one person. The person performs a random walk biased towards the single exit in the room. For example, in Fig.~\ref{fig:latticeGasMovement}(a), the probability that the person takes a step in the $+y$ direction is $P_y = (1 - D)/3 + D e_y /(e_x + e_y)$, while the probability that it takes a step in the $-x$ direction is $P_{-x} = (1 - D)/3 + D e_x /(e_x + e_y)$, since the intended direction $\vec{e}$ has positive projections along the $+y$ and $-x$ directions. Along the $+x$ direction, for which $\vec{e}$'s projection is negative, the probability that the person takes a step in the $+x$ direction is $P_x = (1 - D)/3$. Here we see that the random walk probabilities contain an unbiased component $(1 - D)/n$, which is the same for all $n$ permissible directions (in the Tajima-Nagatani model, the $-y$ direction is not permitted), as well as a biased component $D = 0.7$ that favours movement towards the exit.

\subsection{Social force model}
\label{SocialForceModel}
The social force model is one of the most popular models for motion planning in crowds~\cite{Kamphuis:2004uu,Xi:2010uc,Peng:2009cc}. This model is based on the idea that pedestrians move in response to fictitious attractive or repulsive \emph{social forces} produced by obstacles and other pedestrians.  Over the years, several extensions have been made to the social forces model like the modeling of grouping behavior~\cite{Kamphuis:2004uu}. In this paper, we use the Helbing-Moln\'ar-Farkas-Vicsek (HMFV) social force model~\cite{PhysRevE.51.4282}, which is tweaked from the original model introduced by Helbing and Molnar in 1995. In the original model
\begin{equation} \label{eqn:SFeqn}
m_i \frac{d\vec{v}_i}{dt}=\vec{f}_{i}+\sum_{j(\neq i)}\vec{f}_{ij}+\sum_{W}\vec{f}_{iW}
\end{equation}
proposed by Helbing and Molnar~\cite{PhysRevE.51.4282}, three types of forces act on a given agent $i$ with mass $m_i$, instantaneous position $\vec{r}_i(t)$, and instantaneous velocity $\vec{v}_i(t)$. The first is a restoring force
\begin{equation}
\vec{f}_i = -m_i \frac{\vec{v}_i-\vec{v}_0}{\tau_i}
\end{equation}
that steers the agent towards the desired velocity $\vec{v}_0$ at a rate determined by the characteristic time $\tau_i = 1$. Here
\begin{equation}
\vec{v}_0=(1-p)V_0\vec{e}_i(t)+p\left<\vec{v}_j\right>_i,
\end{equation}
where $V_0$ is the preferred speed, $\vec{e}_i$ is a vector that points towards the exit, and $(1 - p)$ is the weight given to this desired velocity. With a weight of $p$, agent $i$ also adapts to the average velocity $\left<\vec{v}_j\right>_i$ in its neighborhood. When $p$ is small, agent $i$ moves more along its intended direction $\vec{e}_i(t)$, whereas if $p$ is large, agent $i$ tends to follow where its neighbors are going. We can therefore tune $p$ from $p \approx 0$ (self-directed normal egress) to $p \approx 1$ (panic-driven herding during emergency evacuations). In this paper, we used $p = 0.2$ to simulate an emergency evacuation situation.

The second is a repulsive force
\begin{equation}\label{eqn:repulhuman}
\begin{split}
\vec{f}_{ij} &= \lbrace A e^{(R_{ij} - d_{ij})/B} + k \eta(R_{ij}-d_{ij}) \rbrace\vec{n}_{ij} \\
&\qquad+ \kappa \eta(R_{ij}-d_{ij}) \Delta v^t_{ji}\vec{t}_{ij},
\end{split}
\end{equation}
where
\begin{equation}
\eta(x) = \left\{
  \begin{array}{l l}
    x, & \quad \text{if $x \geq 0$};\\
    0, & \quad \text{if $x < 0$}\\
  \end{array} \right.
\end{equation}
that mimics the \emph{psychological tendency} of agents $i$ and $j$ to move away from each other if they are too close. Here $R_{ij} = R_i + R_j$ is the sum of radii of the two agents, $d_{ij} = |\vec{r}_i - \vec{r}_j|$ is the physical distance between the two agents, and $\vec{n}_{ij} = (n^1_{ij},n^2_{ij}) = (\vec{r}_i-\vec{r}_j)/d_{ij}$ is the unit vector pointing from agent $j$ to agent $i$. Additionally, $A = 2000$ N and $B = 0.08$ m are the repulsion coefficient and the fall-off length of interacting agents respectively~\cite{Helbing:2000ef}. Helbing and Molnar also found it necessary to introduce two other terms in the interaction force when agents $i$ and $j$ are in contact with each other, i.e. $d_{ij} < R_{ij}$. This counteracting body compression term $k \eta(R_{ij}-d_{ij})\vec{n}_{ij}$ and sliding friction term $\kappa \eta(r_{ij}-d_{ij})\Delta v^t_{ji}\vec{t}_{ij}$ are crucial for getting realistic behaviors of panicking crowds. Here $\vec{t}_{ij} = (-n^2_{ij},n^1_{ij})$ is the tangential direction and $\Delta v^t_{ji} = (\vec{v}_j-\vec{v}_i)\cdot\vec{t}_{ij}$ is the tangential velocity difference.

Finally, in the third force
\begin{equation}\label{eqn:repulwall}
\begin{split}
\vec{f}_{iW} &= \left\{ A_i e^{(R_{i}-d_{iW})/B_i}+ k \eta(R_i - d_{iW}) \right\}\vec{n}_{iW} \\
&\qquad - \kappa \eta(r_i-d_{iW})(\vec{v}_i\cdot\vec{t}_{iW})\vec{t}_{iW},
\end{split}
\end{equation}
the first and second terms repels agent $i$ from a wall that it is $d_{iW}$ away from, while the third term (which is negative) is introduced to mimic the observation that people move faster near walls when they are in crowded situations. Here, $\vec{n}_{iW}$ is the normal vector of the wall, and $\vec{t}_{iW}$ is the tangent vector of the wall. Also, $k=12,000$ kg/s$^{2}$ and $\kappa=24,000$ kg/ms are respectively the body force constant and the sliding friction force constant used.

\subsection{Reciprocal velocity obstacles model}
\label{RVOModel}

The third model we consider is the reciprocal velocity obstacles (RVO) model. In this model, the time to the next collision is calculated based on the relative velocities of $N$ agents. Each agent then changes its velocity $\{\vec{v}_i\}_{i=1}^N$ to maximize this time to collision. By continuously updating the velocities in this manner, collisions are avoided. This algorithm was first proposed by Fiorini et al.~~\cite{Fiorini:1993hi}, but was first used in multi agent systems in~\cite{vandenBerg:2008cq}. Since then there have been several modifications and improvements to the algorithm~\cite{Guy:2009gu,Guy:2010ko,Guy:2010te,Guy:2010uv}, although the underlying idea remained the same. In this paper, we use the RVO2 model introduced by Guy et al.~\cite{Guy:2010ko}, where the collision avoidance computation is based on computational geometry and linear programming.

Given a preferred velocity, RVO2 helps an agent find the velocity closest to the preferred velocity that will enable it to avoid collisions with all other agents. This is done by determining for each neighboring dynamic and static obstacle an \emph{Optimal Reciprocal Collision Avoidance line (ORCA line) }. Each ORCA line determines the half plane in the velocity plane where the agent's velocity should lie to ensure that no collision occurs with that particular obstacle for the next $\tau$ seconds. $\tau$ is a parameter that specifies the number of seconds for which the chosen velocity should avoid a collision. From the set of half planes thus obtained, the optimal collision-avoiding velocity can be determined as the velocity within the permissible region for all the half planes that is closest to the preferred velocity. This can be determined efficiently using linear programming. Besides the speed and efficiency of the algorithm, another strong appeal of this model is that we need only set one parameter $\tau$. For the experiments in this paper we use two values of $\tau$;  0.5 seconds or 10 simulation time steps for avoiding dynamic agents and  0.05 seconds or 1 simulation time step for avoiding static obstacles. The value of $\tau$ translates in practical terms to how early a walker tries to avoid a collision with an obstacle.

\section{Methodology}
\label{Methodology}

\begin{figure}[htbp]
\includegraphics[width=3.4in]{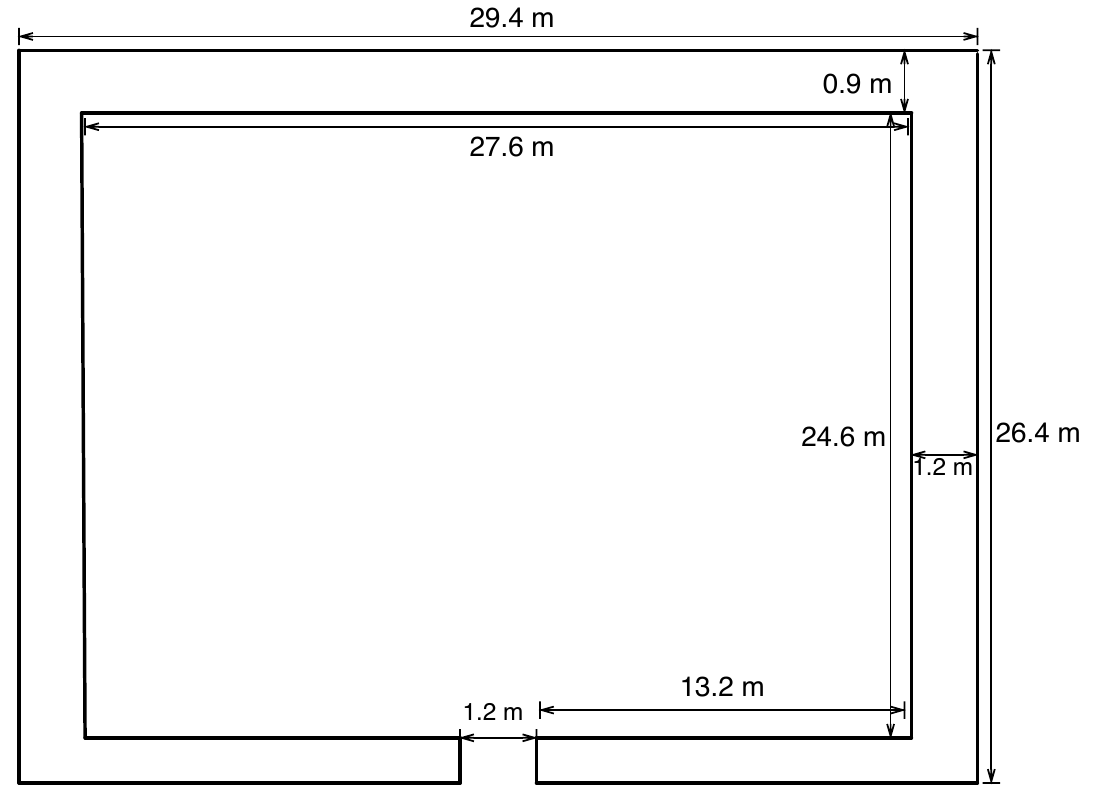}
\caption{The environment setup used for the simulations in this paper.}
\label{fig:experimentalSetup}
\end{figure}

In this section we explain the simulations that were carried out and the analysis done. Java 6 and the MASON simulation framework~\cite{Luke:2005wc} were used for creating and running the simulations. The simulated scenario consists of agents evacuating from a simple rectangular room with a single exit. The dimensions of the room are shown in Fig.~\ref{fig:experimentalSetup}. Evacuation of the room was simulated with $N = 50$, 100, 150, 200, 300, 500 and 1000 agents whose locations were randomly distributed within the room. For each $N$, we ran 100 simulations for each model, and the positions and velocities of all agents at all time steps were collected. For meaningful comparison between the three models, we assumed that each agent is a perfect circle with radius $R = 15$ cm~\cite{Pan:2006vp}, that has a preferred speed of 1.3 ms$^{-1}$, and a maximum speed of 2.6 ms$^{-1}$~\cite{fruin1992designing}. For the social force model we also assumed that all agents have the same mass of 60 kg.

Where possible, we have used the same settings for all three models to ensure a fair comparison. The values used for the remaining parameters required for the models was given in Sec.~\ref{Models} along with the respective models. However, to the best of our knowledge, only the social force model has been calibrated against real world data~\cite{Bauer2011}. The other two models are not calibrated beyond choosing a reasonable and believable model and the values chosen are taken such that they produce empirically believable results.

\subsection{Measurements}

Depending on the context and motivation of the model there are several metrics that are generally used. For CA models that are generally used for studying macroscopic patterns, it is common to measure macro variables like the mean evacuation time~\cite{Nagai:2004kl} and flow rate~\cite{Tajima:2001to}. For models like social force where the motivation is to study interactions at a more granular level, it is common to make empirical and qualitative observations like the lane effect~\cite{PhysRevE.51.4282}. Sometimes quantitative metrics like density-dependence of the flow or velocity are also measured~\cite{Seyfried2008}. From a graphics perspective, where performance is key, computation time~\cite{Ondrej:2010hv}, frame rate and run time per frame are generally measured against the number of walkers~\cite{vandenBerg:2008fu}.

\begin{figure}[h]
\begin{center}
\includegraphics[scale=0.5]{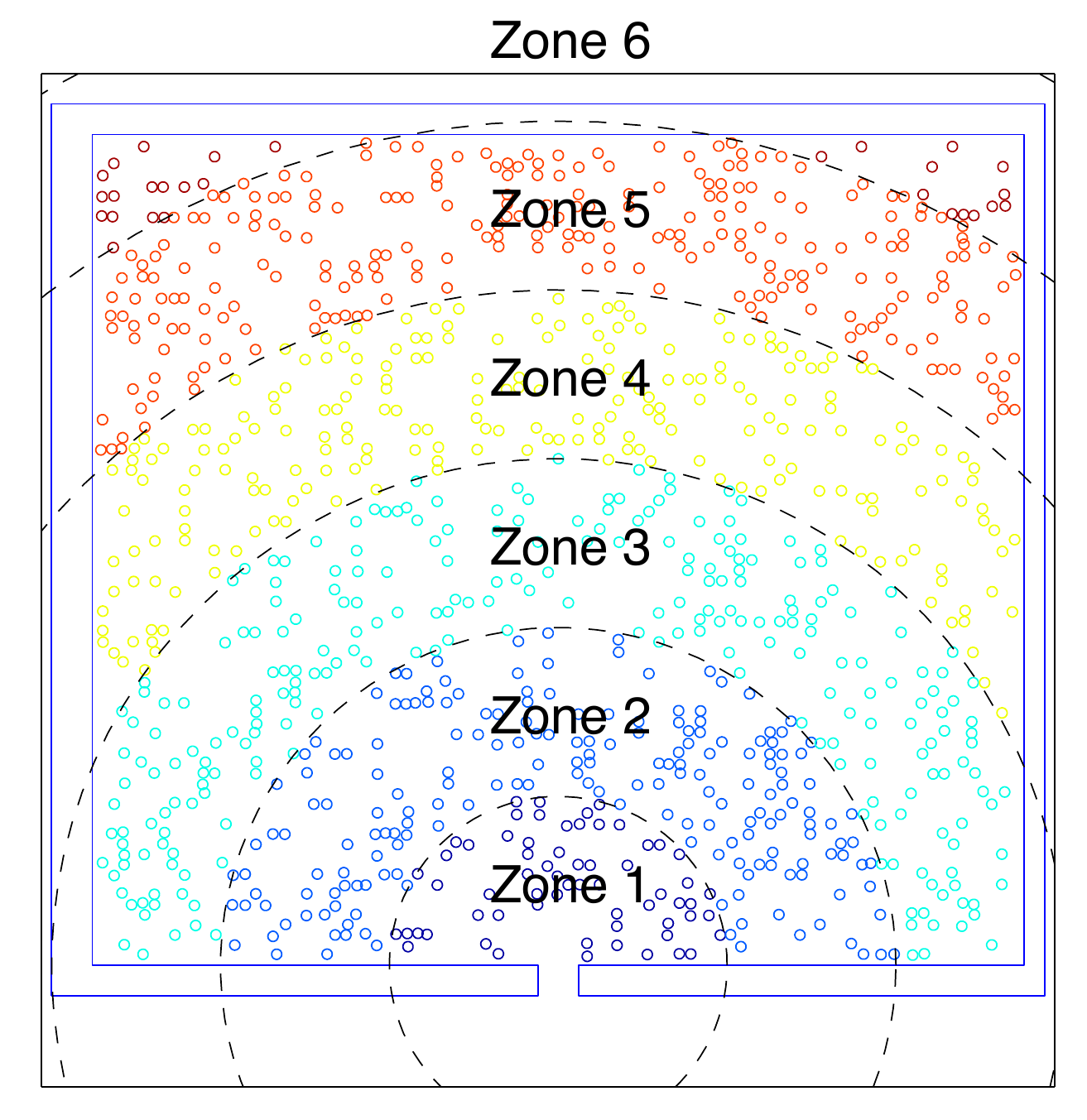}
\caption{In our simulations, agents are grouped into six zones based on their initial distances from the exit. The outer radii of Zones 1, 2, 3, 4, 5 and 6 are 5 m, 10 m, 15 m, 20 m, 25 m and 30 m respectively.}
\label{fig:Zoning}
\end{center}
\end{figure}

In this paper, we performed two main classes of measurements: time-based and distance-based measurements. For time-based measurements, we first measured the evacuation time distributions for different number of agents. To better understand the stages in the evacuation, we also divide the room into six different zones (see Fig.~\ref{fig:Zoning}) to measure the evacuation time sub-distribution. The outer radii of Zones 1, 2, 3, 4, 5 and 6 are 5 m, 10 m, 15 m, 20 m, 25 m and 30 m respectively. We also measure the flow rate at the exit as a function of time for each model.

For distance-based measurements, we traced the trajectories of all agents to obtain the distribution of total distance travelled going from the initial position to the exit. Here, the total distance travelled $D_i$ by agent $i$ is just the sum of its displacements $\sum^{T_i}_{t=0}||\vec{r}_i(t+1)-\vec{r}_i(t)||$, where $T_i$ is the evacuated time of agent $i$. We also divide the room into $100 \times 100$ cells, and count the number of agents passing through each cell as the simulation progresses. This is then visualized as a heat map. Finally, we measure the ratio of the total distance travelled $D_i$ by agent $i$ to the minimum distance $D_{\min}$ it would cover if it evacuated along a straight line (For lattice model, $D_{min}$ is the Manhattan distance between initial position of agent and exit). This in some sense quantifies the `inconvenience', $I_i$ experienced by the agent $i$ during the evacuation.
\begin{equation}\label{eqn:Incon}
I_i=\frac{D_i}{D_{\min}}
\end{equation}

\subsection{DISTATIS}\label{sec:distatis}

After making six different measurements on three models, we discovered that when we look at the flow rates, the lattice gas model is more similar to the RVO2 model. However, if we look at the total distance travelled, the social force model is more similar to the RVO2 model. If we look at the zoned evacuation times, we find that the three models have very different distributions (see Sec.~\ref{EvacTime}). Since we do not know \emph{a priori} which measurements best discriminate the three models (or for that matter, best discriminate models from the real world), we want to be able to compare the three models quantitatively, incorporating information from all six measurements.

To do so, we compute the \emph{Jensen-Shannon divergence}~\cite{Lin:1991it}
\begin{equation}
D_{JS}[f_{k\mu}, f_{k\nu}] = H[f_k] - \frac{1}{2} H[f_{k\mu}] - \frac{1}{2} H[f_{k\nu}] \geq 0
\end{equation}
between the distributions $f_{k\mu}(z)$ and $f_{k\nu}(z)$ for measurement $k$ of models $\mu$ and $\nu$. Here, $z$ is a continuous variable like evacuation time, distance travelled, or inconvenience,
\begin{equation}
H[f] = -\int dz\, f(z) \ln f(z)
\end{equation}
is the \emph{Shannon information function}, and
\begin{equation}
f_k(z) = \frac{1}{2}\left[f_{k\mu}(z) + f_{k\nu}(z)\right].
\end{equation}
If $f_{k\mu}(z)$ and $f_{k\nu}(z)$ are highly similar, we will get $D_{JS}[f_{k\mu}, f_{k\nu}] \approx 0$, whereas if $f_{k\mu}(z)$ and $f_{k\nu}(z)$ are very different, $D_{JS}[f_{k\mu}, f_{k\nu}] \gg 0$, i.e. the Jensen-Shannon divergence qualifies as a distance metric. In this way, we obtain six $3 \times 3$ distance matrices $\vec{D}_k$.

\begin{figure}[htbp]
\centering
\includegraphics[width=\linewidth]{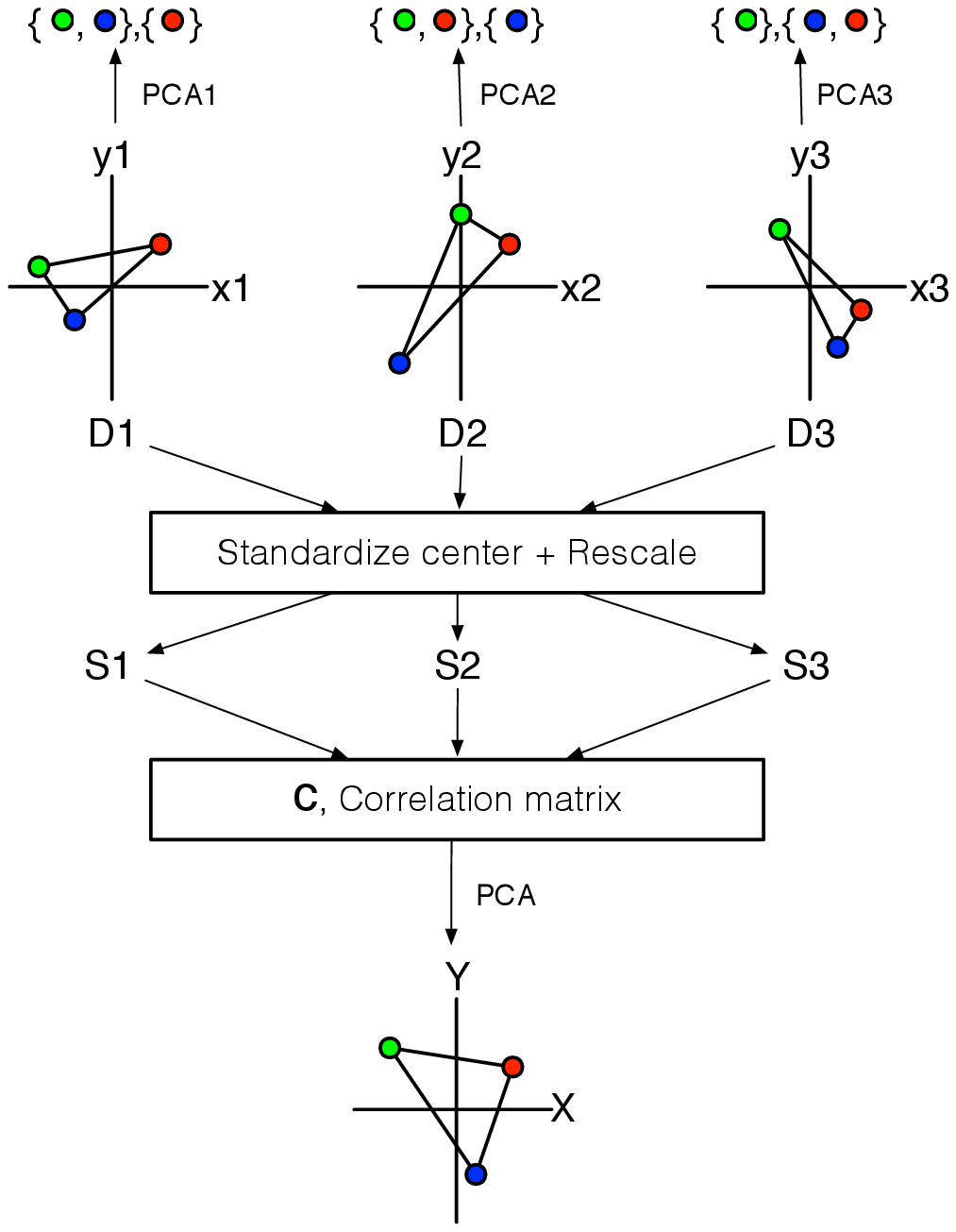}
\caption{In many classification problems, we can define different distance metrics for the same set of data points to measure how dissimilar they are to each other. In this figure we show three different distance matrices D1, D2, and D3. If we perform principal component analysis (PCA) on them separately (PCA1, PCA2, PCA3), we are not likely to agree on which data points are more similar to each other. In the DISTATIS method, we first standardize the distance matrices by centering and rescaling, to get S1, S2, and S3. If two data points are indeed more similar to each other than they are with the third, then we expect this similarity structure to be reflected in S1, S2, and S3, i.e.~two standardized distance matrices would have very similar patterns of matrix elements. With this in mind, we reshape the matrices S1, S2, and S3 to make them into vectors of distances. We then compute the cross correlations between these vectors, before performing PCA on the correlation matrix C. This PCA tells us which distance matrices agree with each other more, and which less. The result is a consensus between different distance metrics, based on which a single PCA will discover the similarity structure between the data points.}
\label{fig:DISTATIS}
\end{figure}

Then, we use the \emph{DISTATIS} method~\cite{Abdi:DISTATIS} for analyzing multiple distance matrices. This is a generalization of the method of principal component analysis (PCA). STATIS is an acronym for the French expression `Structuration des Tableaux \`{a} Trois Indices de la Statistique', which roughly means `structuring three way statistical tables'. The difference between a straightforward application of the PCA method and the DISTATIS method is shown in Fig.~\ref{fig:DISTATIS}. Instead of six different PCAs for the distance matrices obtained from the six different variables measured, and inevitably end up with conflicting conclusions on which models are more similar, we standardize the six distance matrices, and ask which variables are more similar to each other.

The idea behind DISTATIS is that the data points are similarly clustered in two variables, if their standardized distance matrices in these two variables are similar to each other. Therefore, if we compute the cross correlations between variables, we will discover which variables give more similar outcomes to which other variables through a PCA. Components of the first eigenvector obtained from this PCA tells us how important each variable is. By weighting the distance matrices of each variable by its component in the first eigenvector, we construct a \emph{compromise matrix} whose matrix elements give us the most reliable similarity/dissimilarity between data points. A final PCA of this compromise matrix then gives the most reliable groups of data points. In Fig.~\ref{fig:Distatis1000}, we show the six independent PCA results superimposed on the PCA of the compromise matrix.

\section{Results and Discussion}
\label{Discussion}

The six measurements on the simulations are the (1) evacuation time, (2) zoned evacuation time, (3) passage density map, (4) total distance traveled, (5) inconvenience and (6) flow rate.

Here we show the results for (1) evacuation time (Fig.~\ref{fig:EvacTime}) and (2) zoned evacuation time (Fig.~\ref{fig:ZoningEvac}), (3) passage density map (Fig.~\ref{fig:DensityMap}), (4) total distance traveled (Fig.~\ref{fig:TravelDist}), (5) inconvenience (Fig.~\ref{fig:Incon}) and finally flow rate (Fig.~\ref{fig:FlowRate}) for all three models, with different number of agents.

\subsection{Evacuation time}
\label{EvacTime}

\begin{figure*}[htbp]
\subfigure[]{\includegraphics[width=6cm]{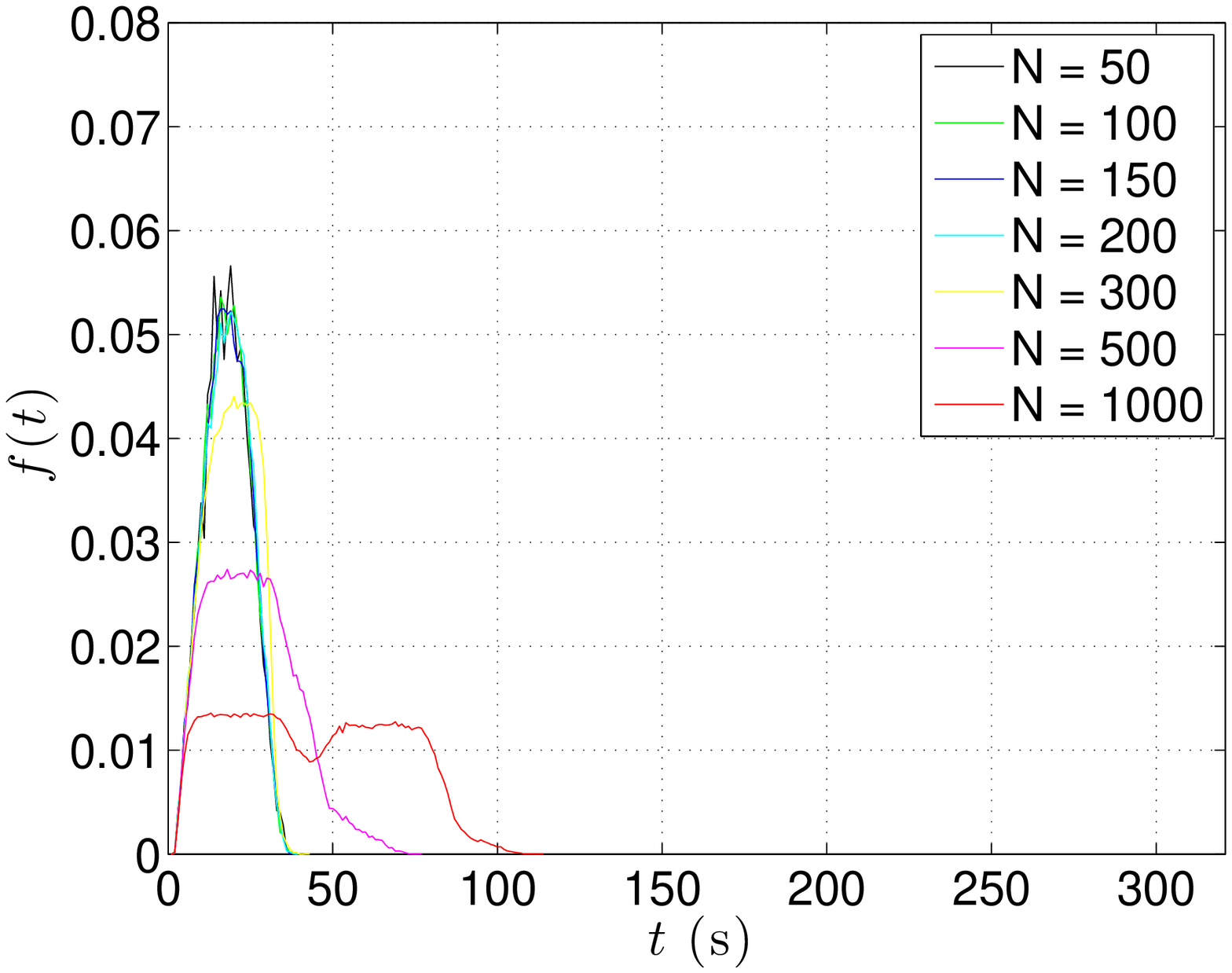}\label{fig:LatticeEvacTime}}
\subfigure[]{\includegraphics[width=6cm]{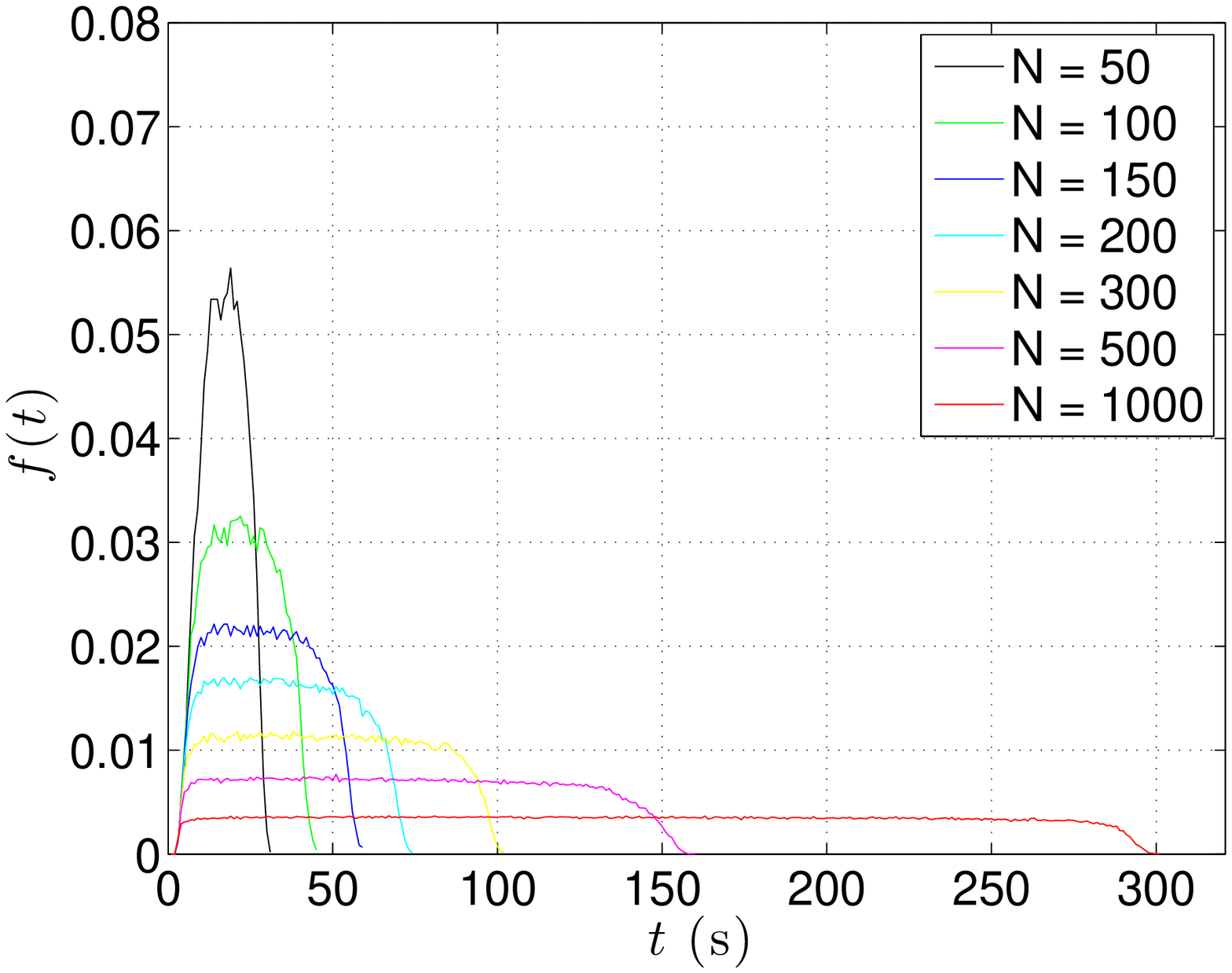}\label{fig:SocialForceEvacTime}}
\subfigure[]{\includegraphics[width=6cm]{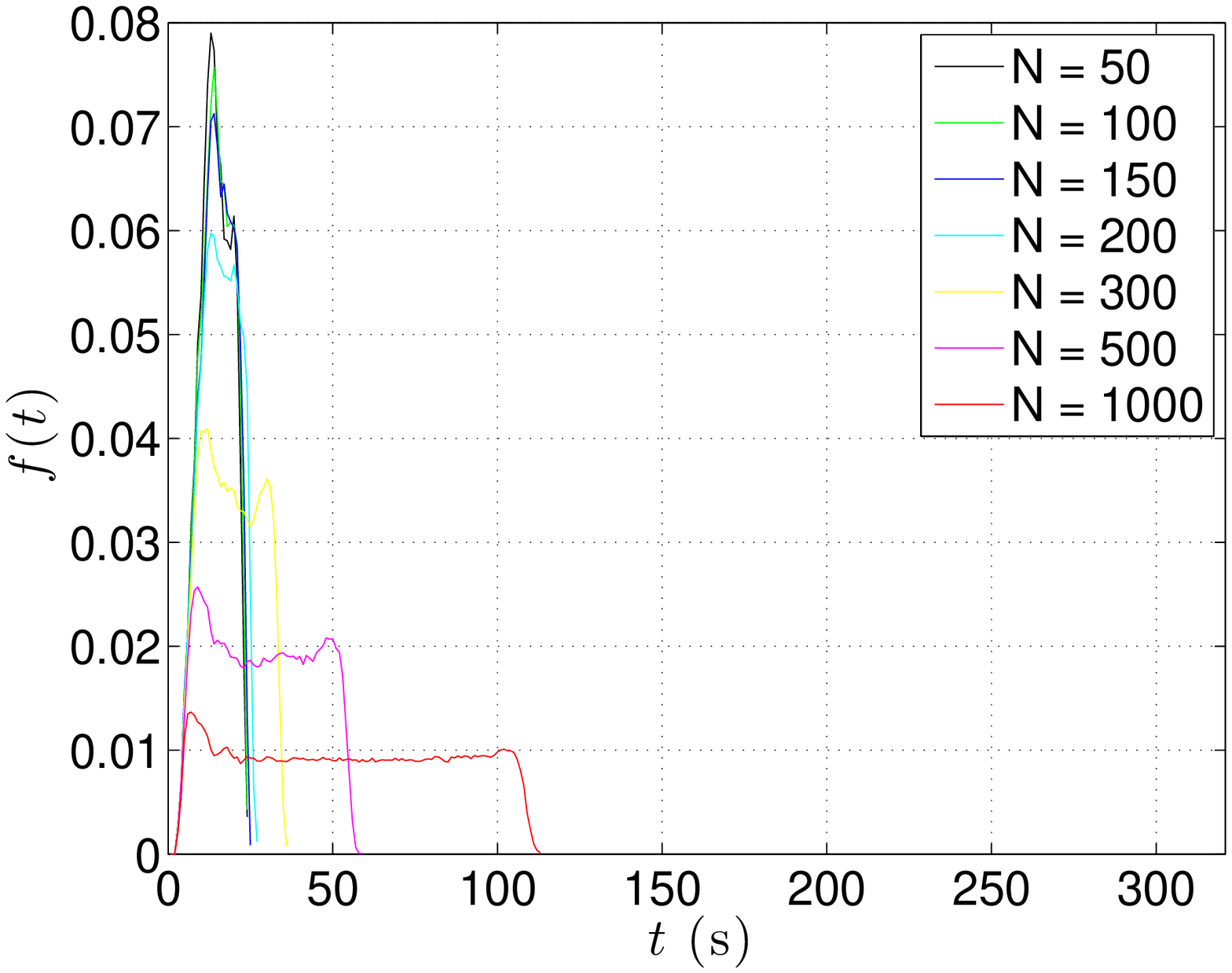}\label{fig:RVO2EvacTime}}
\caption{Evacuation time distributions for the (a) lattice gas, (b) social force, and (c) RVO2 models for $N = 50, 100, 150, 200, 300, 500, 1000$ agents. For each model and $N$, the distribution is built from 100 simulations.}
\label{fig:EvacTime}
\end{figure*}

As Fig.~\ref{fig:EvacTime} shows, the evacuation time distributions for all models have a uniform sub-distribution in the middle of the evacuation, which signals congestion at the exit, once we have $N > 50$ agents. However, in the RVO2 model this uniform sub-distribution is flanked by two small peaks. These two small peaks suggest a higher evacuation rate \emph{just before} congestion sets in and \emph{just before} congestion dissipates in the RVO2 model. This is not observed in real crowds, and hence is clearly an artifact of the model.

Comparing the three models, we find that the social force model has the longest evacuation times of up to five minutes. More importantly, careful inspection reveals fluid-like crowd dynamics around the exit in the RVO2 and lattice gas models. These can be seen in our  videos \url{http://www.youtube.com/watch?v=P64p3nlH_P4} and  \url{http://www.youtube.com/watch?v=vJ0Nzi5Bykw}. In the RVO2 model, an obvious back flow can also be seen during the sudden gush of agents toward the exit. Due to the strong repulsion forces, we see a solid-like behavior in the social force model when the exit becomes congested (see our video \url{http://www.youtube.com/watch?v=3Ot7m959_yo}).

\begin{figure*}[htbp]
\subfigure[]{\includegraphics[width=6cm]{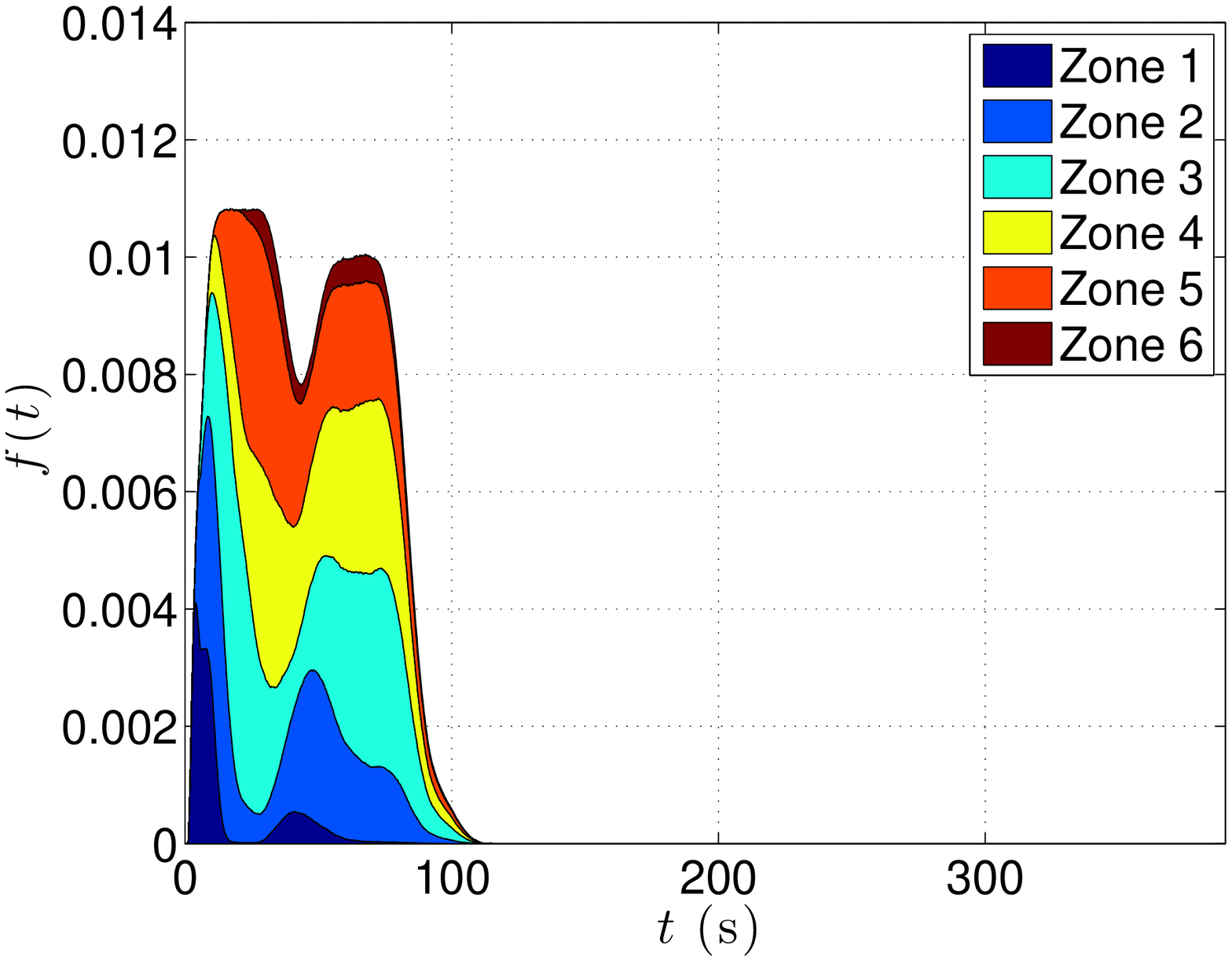}\label{fig:LatticeZoningEvac1000}}
\subfigure[]{\includegraphics[width=6cm]{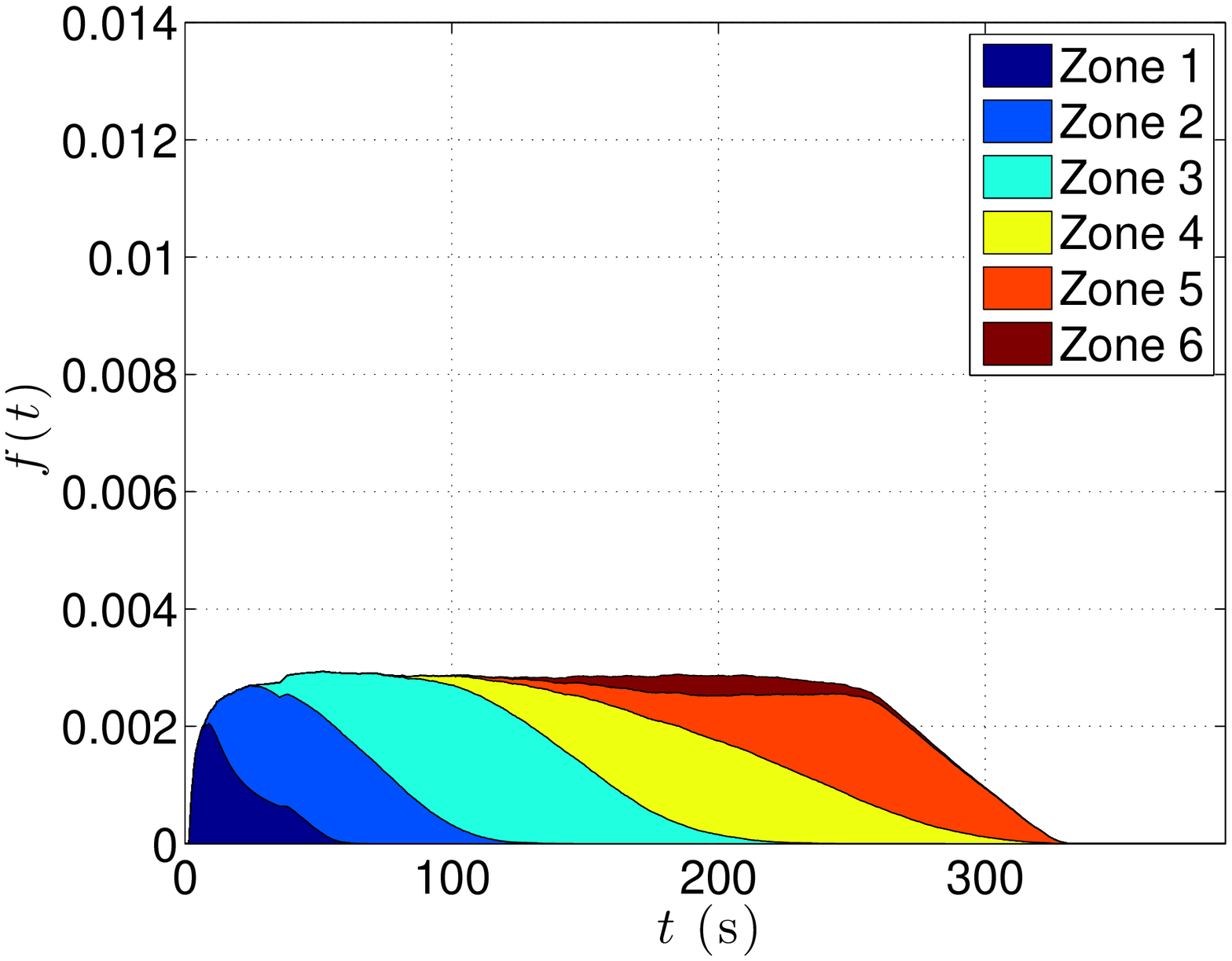}\label{fig:SocialForceZoningEvac1000}}
\subfigure[]{\includegraphics[width=6cm]{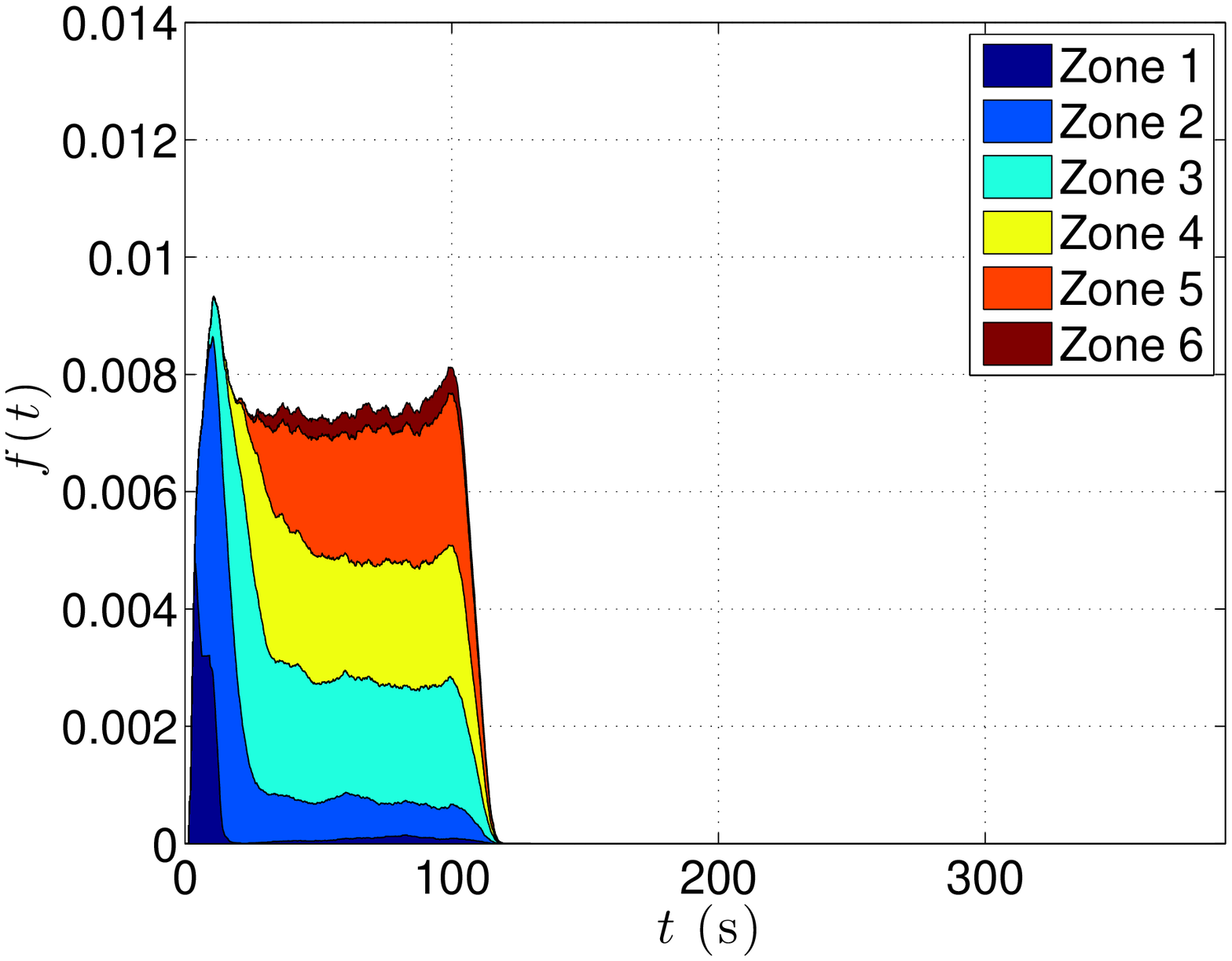}\label{fig:RVO2ZoningEvac1000}}
\caption{Zoned evacuation time distributions for the (a) lattice gas, (b) social force, and (c) RVO2 models for $N = 1000$ agents in six zones, with zone 1 closest to the exit and zone 6 furthest away (see Figure~\ref{fig:Zoning}). For each model, the distribution is averaged over 100 simulations.}
\label{fig:ZoningEvac}
\end{figure*}

In addition, we divided the room into six zones, as shown in Fig.~\ref{fig:Zoning}, based on the distance to the exit, to determine how strongly the crowds mix in the three models. The distributions of zoned evacuation times are shown in Fig.~\ref{fig:ZoningEvac}. These tell us that mixing is strong in the lattice gas and RVO2 models, but weak in the social force model. This mixing dynamics can be seen more clearly from our videos \url{http://www.youtube.com/watch?v=qeoJotgEUxk} (lattice gas), \url{http://www.youtube.com/watch?v=uZpd5LODYZs} (RVO2), and \url{http://www.youtube.com/watch?v=wEB6Ya0o_yw} (social force).
From the movies of the lattice gas and RVO2 simulations, we see that agents from the nearer zones get pushed to the side once there is congestion at the exit. In these two models, agents prefer to keep moving when it is possible for them to do so. Agents in the social force model behave differently: once the exit becomes congested, they will become nearly stationary and wait for their turn to go through the exit.

\subsection{Passage density}

\begin{figure*}[htbp]
\subfigure[]{\includegraphics[width=6cm]{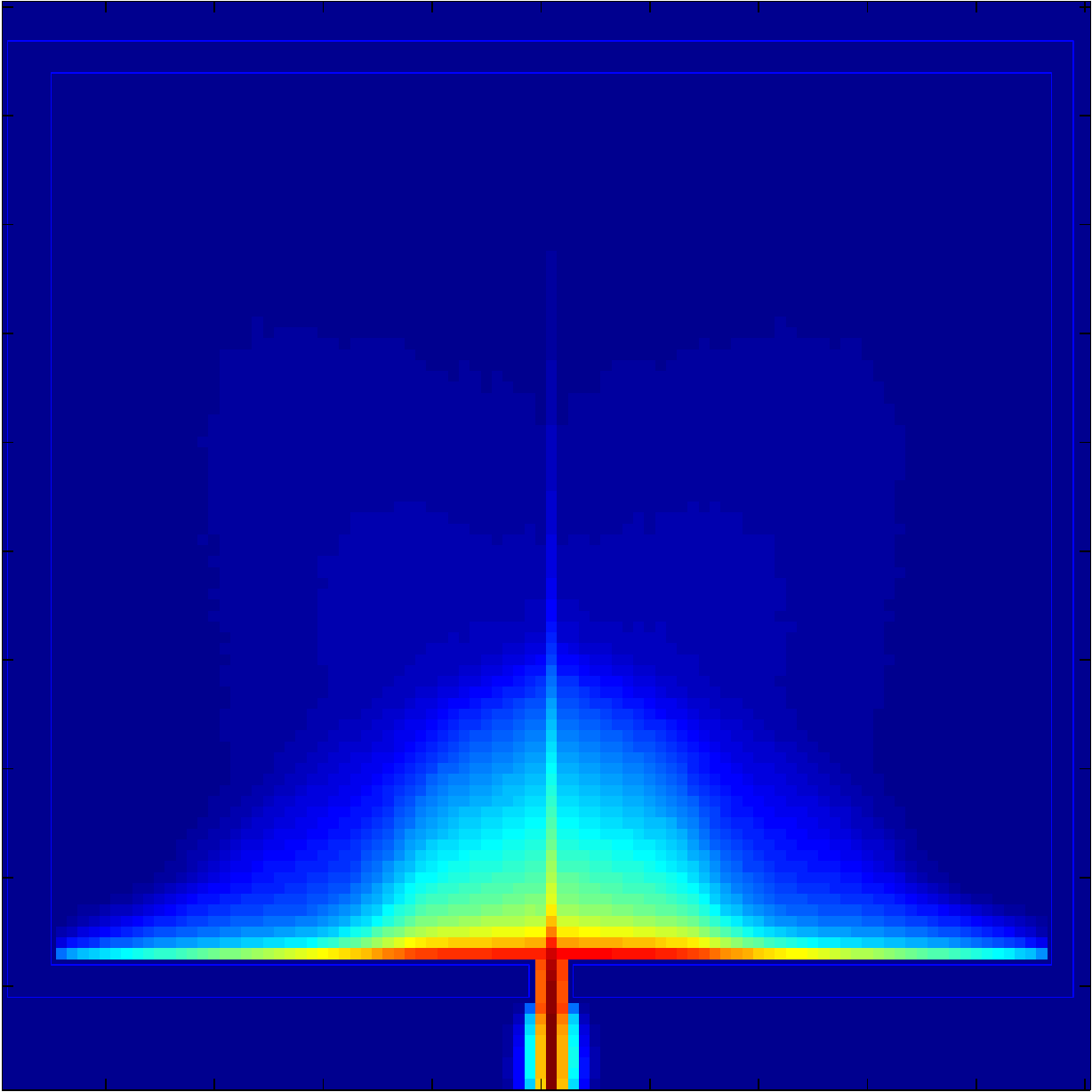}\label{fig:LatticeDensityMap}}
\subfigure[]{\includegraphics[width=6cm]{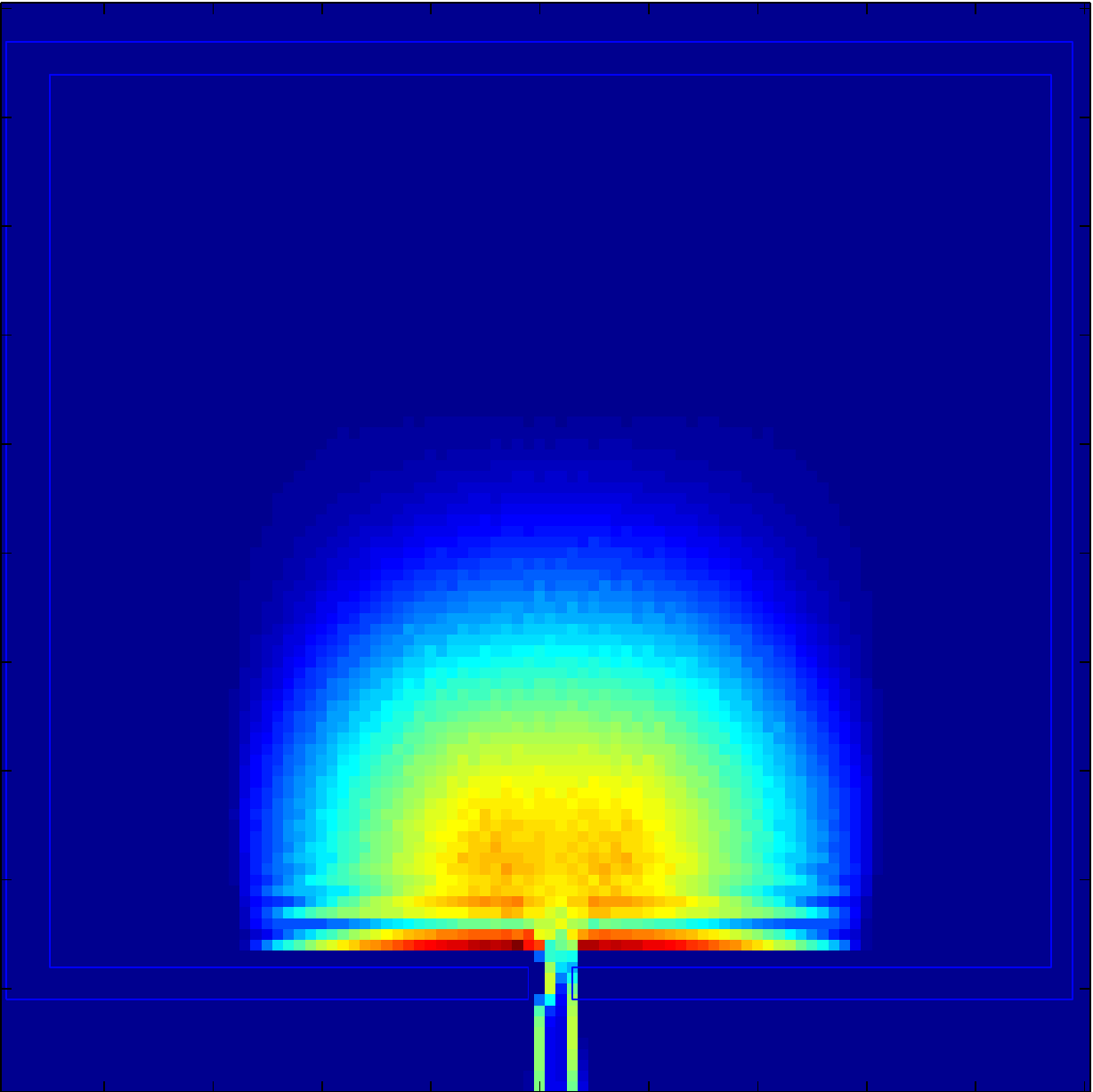}\label{fig:SocialForceDensityMap}}
\subfigure[]{\includegraphics[width=6cm]{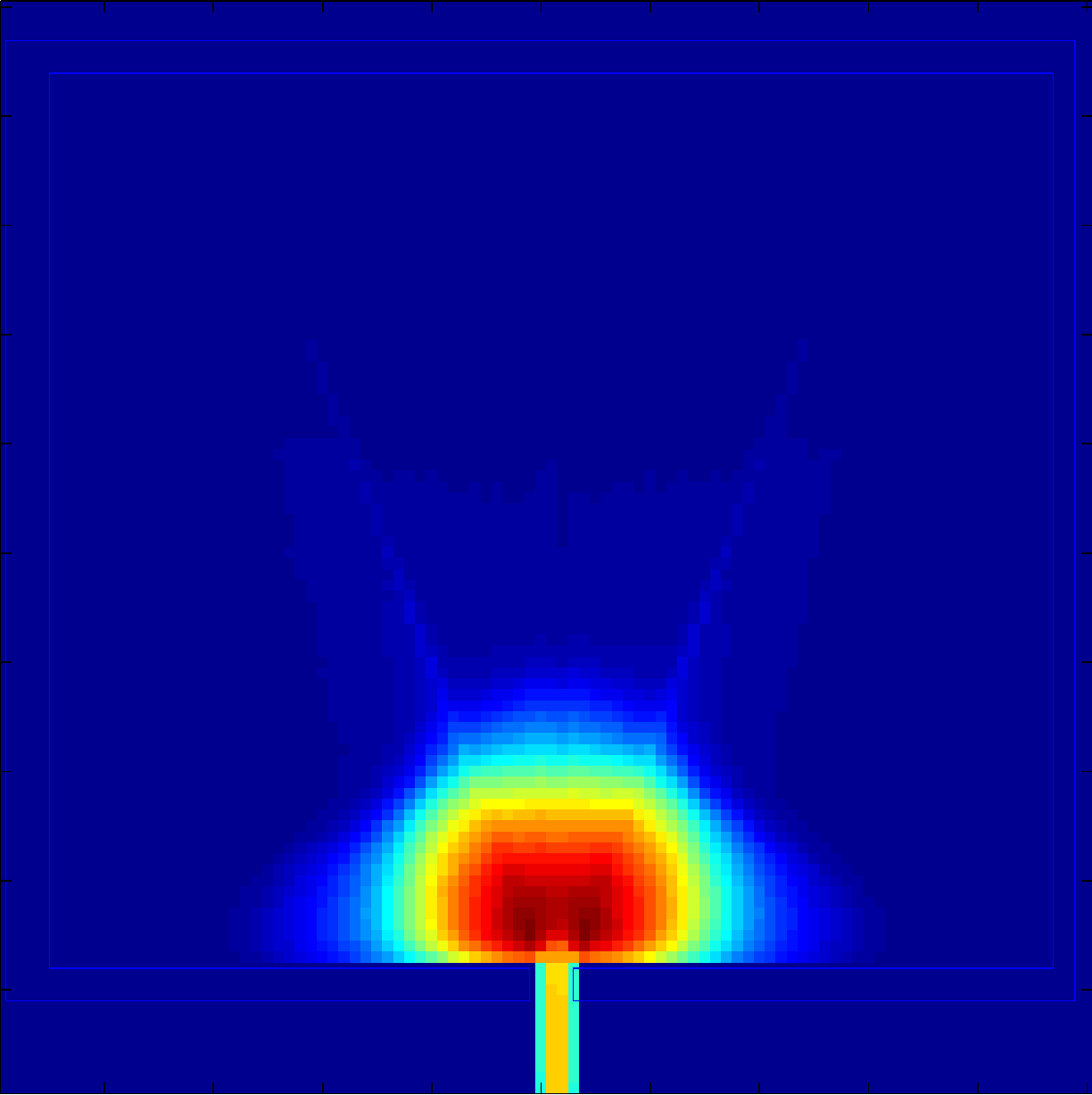}\label{fig:RVO2DensityMap}}
\caption{Passage density maps of $N = 1000$ agents averaged over 100 simulations for the (a) lattice gas, (b) social force, and (c) RVO2 models. In these color maps, the exit is located at the bottom.}
\label{fig:DensityMap}
\end{figure*}

The passage density heat map of agent locations averaged over 100 simulations provides information about the spatial-temporal trace of agents during their evacuation. This is crucial when analyzing the spatial structure of congestions. In Fig.~\ref{fig:LatticeDensityMap}, we see a channel leading straight through the exit. This channel acts as an attractor, because once an agent gets onto this, it will be forced statistically towards the exit ($P_{t,y} = 1.00$). Though not as pronounced as for the lattice gas model, two left-right symmetric channels also formed in the RVO2 model (Fig.~\ref{fig:RVO2DensityMap}). These two channels point roughly from the centers of mass of the left and right halves of the room towards the exit. No such artificial channels can be seen in the passage density map of the social force model (Fig.~\ref{fig:SocialForceDensityMap}).

Comparing the three models, we also see that the RVO2 model (Fig.~\ref{fig:RVO2DensityMap}) has the most compact passage density, whereas the social force model (Fig.~\ref{fig:SocialForceDensityMap}) has the least compact passage density. In particular, in the region just in front of the exit, the RVO2 model gives rise to very high passage density. This high passage density cannot be attributed to agents stepping over a cell once or twice, as in the lattice gas and social force models. In the RVO2 model, this high passage density is produced by repeated visits to the same cell by individual agents that are always moving.


\subsection{Total distance traveled distribution and inconvenience}
\begin{figure*}[htbp]
\subfigure[]{\includegraphics[width=6cm]{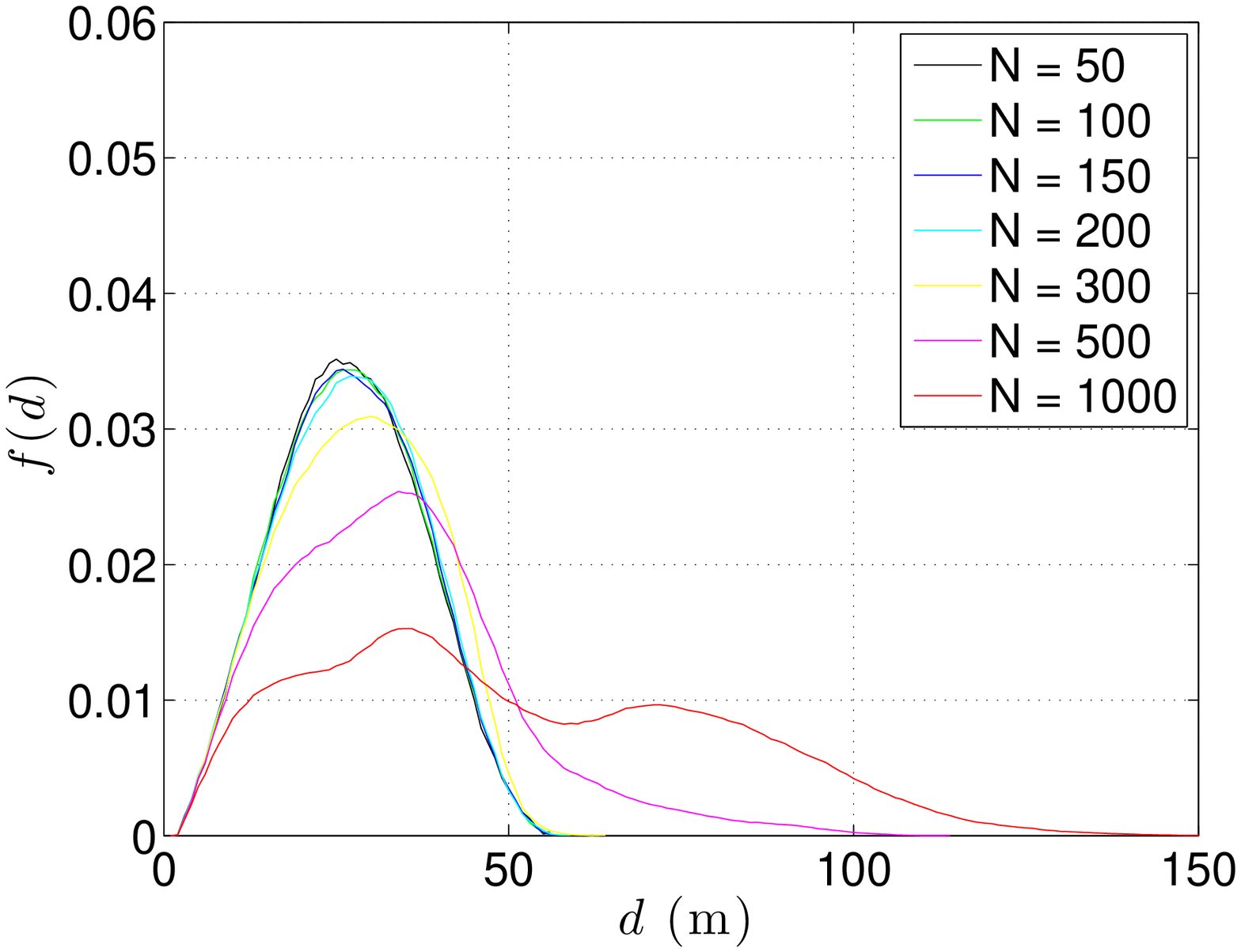}\label{fig:LatticeTravDist}}
\subfigure[]{\includegraphics[width=6cm]{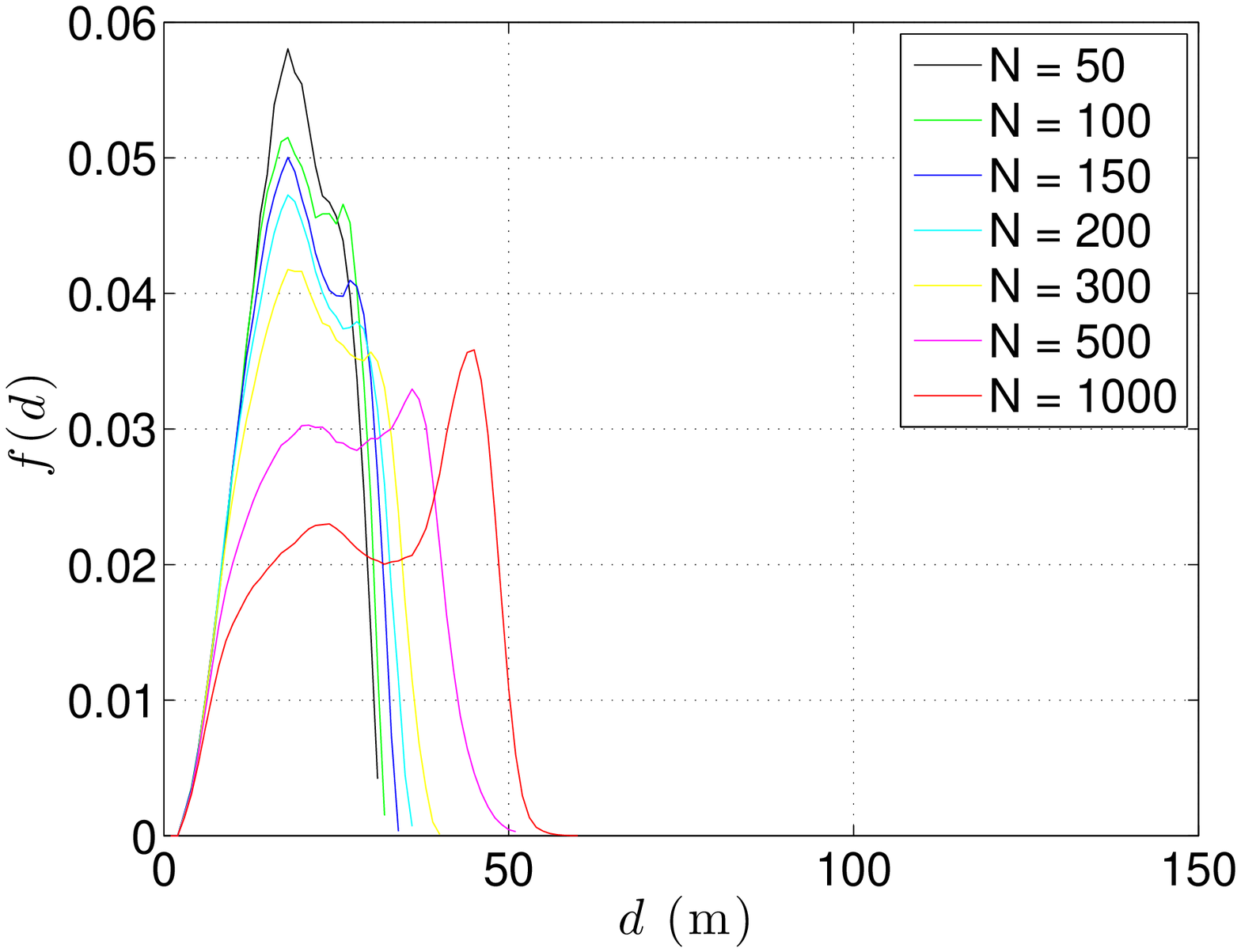}\label{fig:SocialForceTravDist}}
\subfigure[]{\includegraphics[width=6cm]{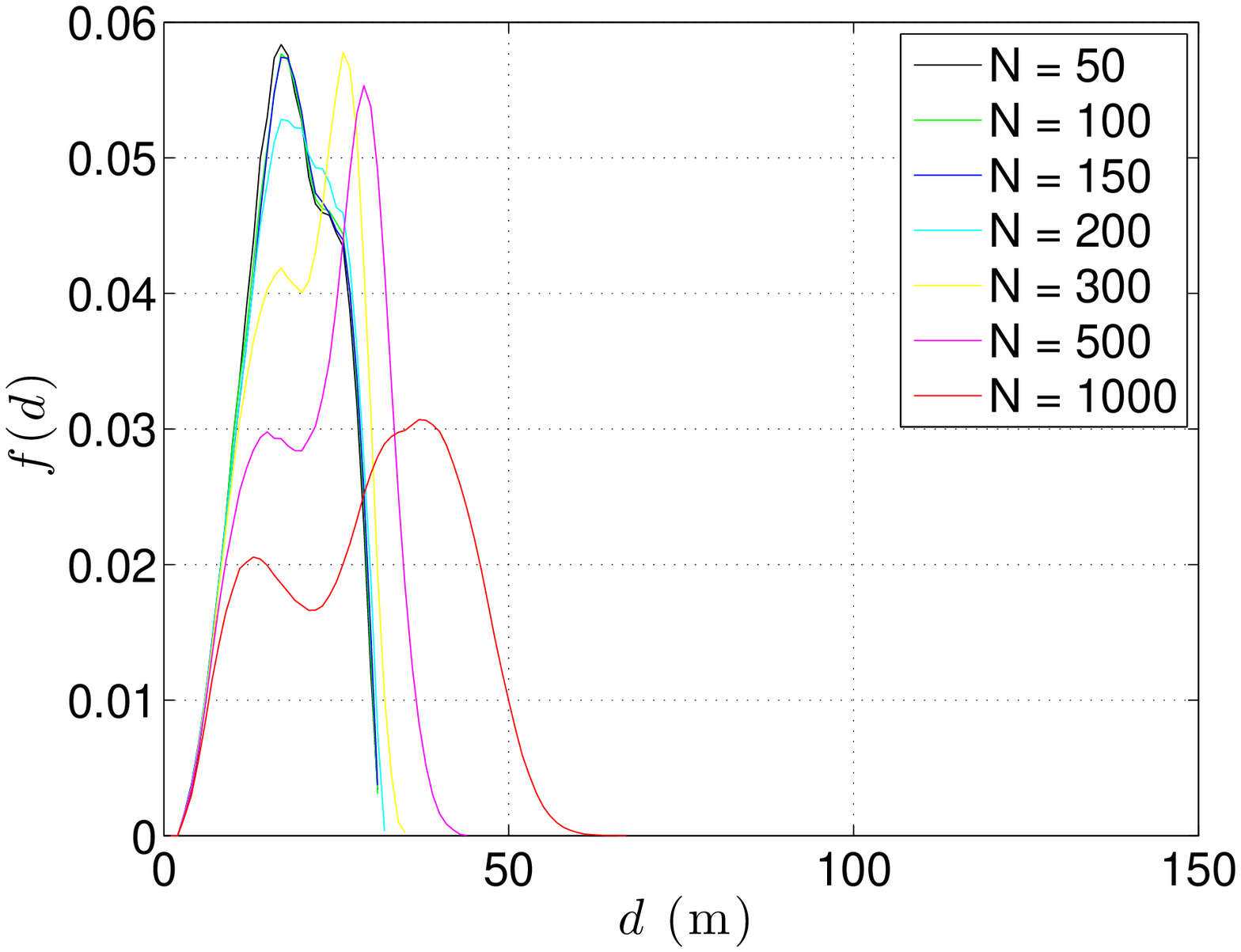}\label{fig:RVO2TravDist}}
\caption{Distributions of total distance traveled for the (a) lattice gas, (b) social force, and (c) RVO2 models. For each model and $N$, the distribution is averaged over 100 simulations.}
\label{fig:TravelDist}
\end{figure*}

To measure the difficulty (inconvenience) for an agent to squeeze through the crowd during evacuation, we measured the \emph{total distance traveled} and as well as the \emph{inconvenience} (defined in Eq.~\ref{eqn:Incon}). As we can see in Fig.~\ref{fig:SocialForceTravDist} and Fig.~\ref{fig:RVO2TravDist}, the distributions of total distance traveled are very similar for the social force and RVO2 models. The lattice gas model, on the other hand, has very different distributions of total distance traveled (Fig.~\ref{fig:LatticeTravDist}). This is because space is continuous in the two former models, but discrete in the lattice gas model.

\begin{figure*}[htbp]
\subfigure[]{\includegraphics[width=6cm]{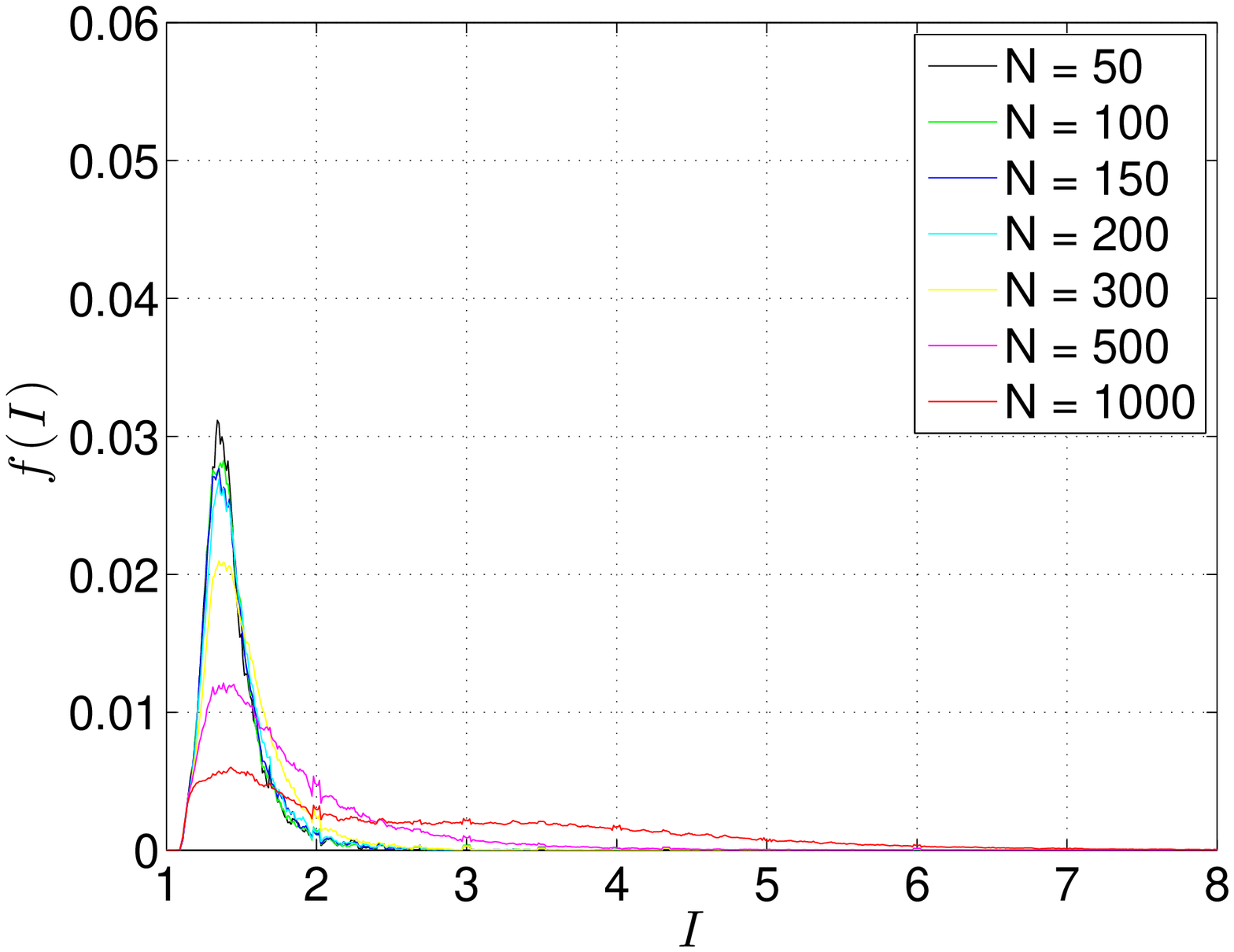}\label{fig:LatticeIncon}}
\subfigure[]{\includegraphics[width=6cm]{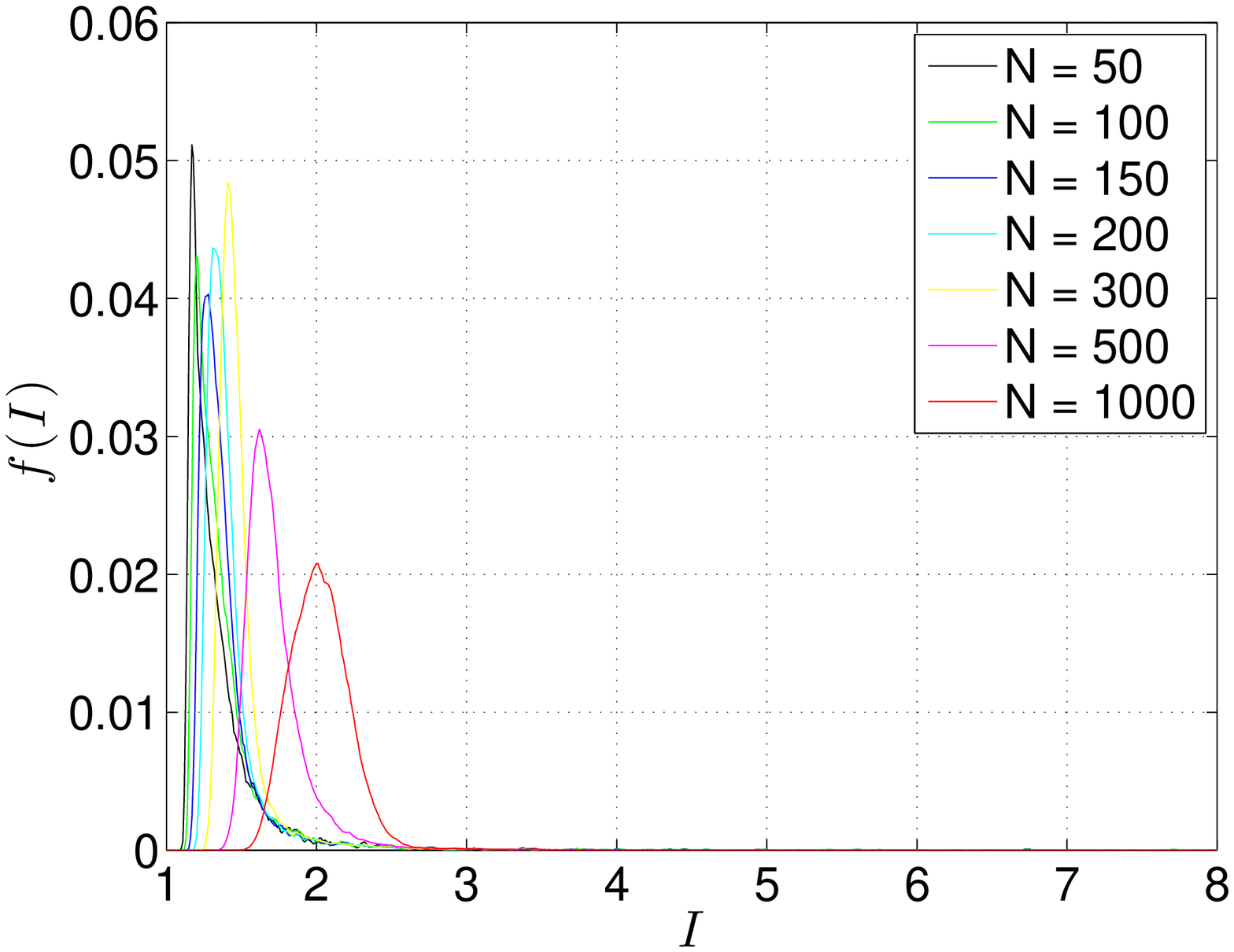}\label{fig:SocialForceIncon}}
\subfigure[]{\includegraphics[width=6cm]{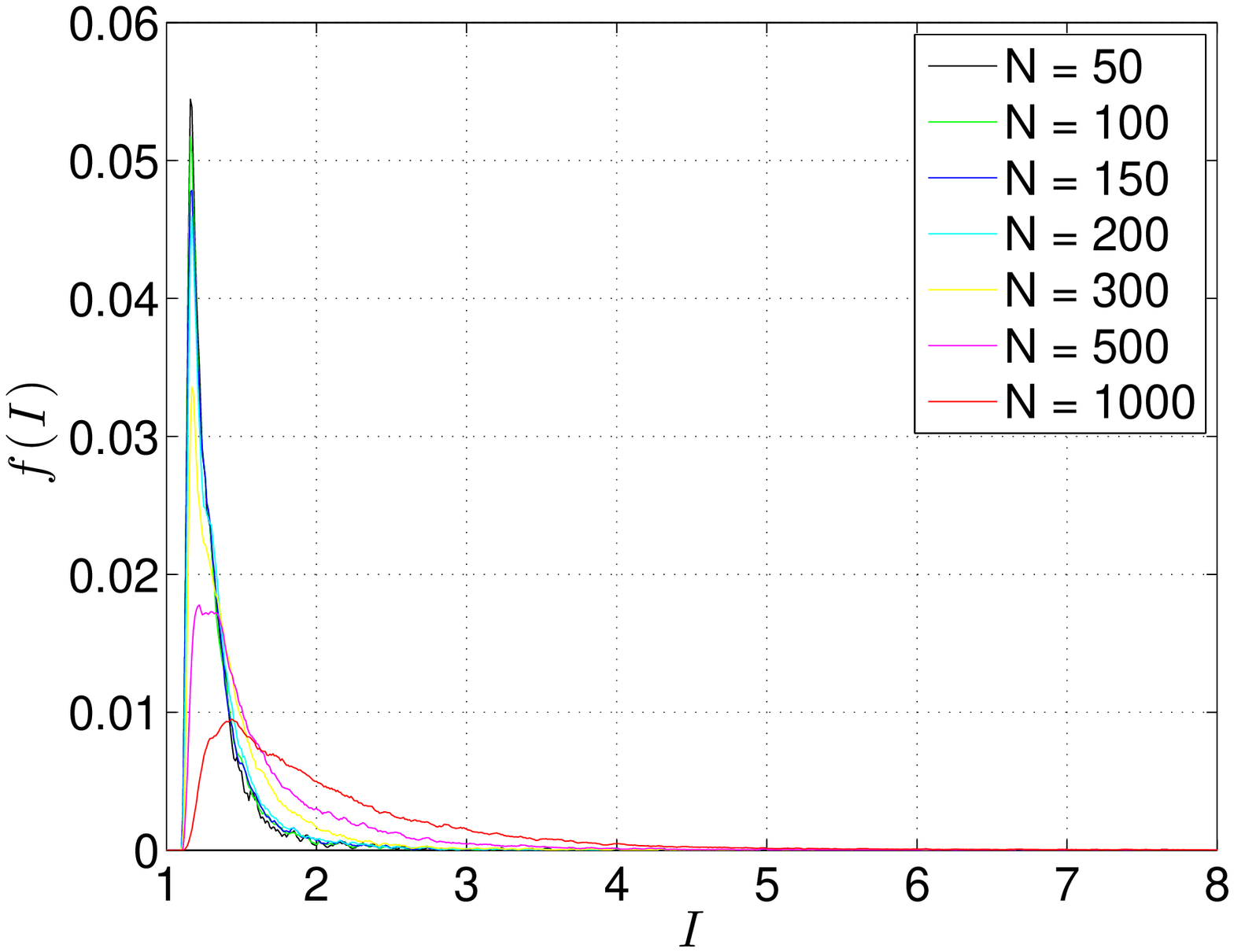}\label{fig:RVO2Incon}}
\caption{Distributions of inconveniences for the (a) lattice gas, (b) social force, and (c) RVO2 models. For each model and $N$, the distribution is averaged over 100 simulations.}
\label{fig:Incon}
\end{figure*}

However, when we compare the distributions of inconveniences in Fig.~\ref{fig:Incon}, we find the lattice gas model is more qualitatively similar to the RVO2 model, though the typical inconveniences in the lattice gas model are larger than the those in the RVO2 model. For different $N$, the distribution peaks around the same inconvenience for both models. In contrast, the peak position of the inconvenience distribution of the social force model depends strongly on the number of agents in the simulation.

\subsection{Evacuation rate}

\begin{figure*}[htbp]

\subfigure[]{\includegraphics[width=6cm]{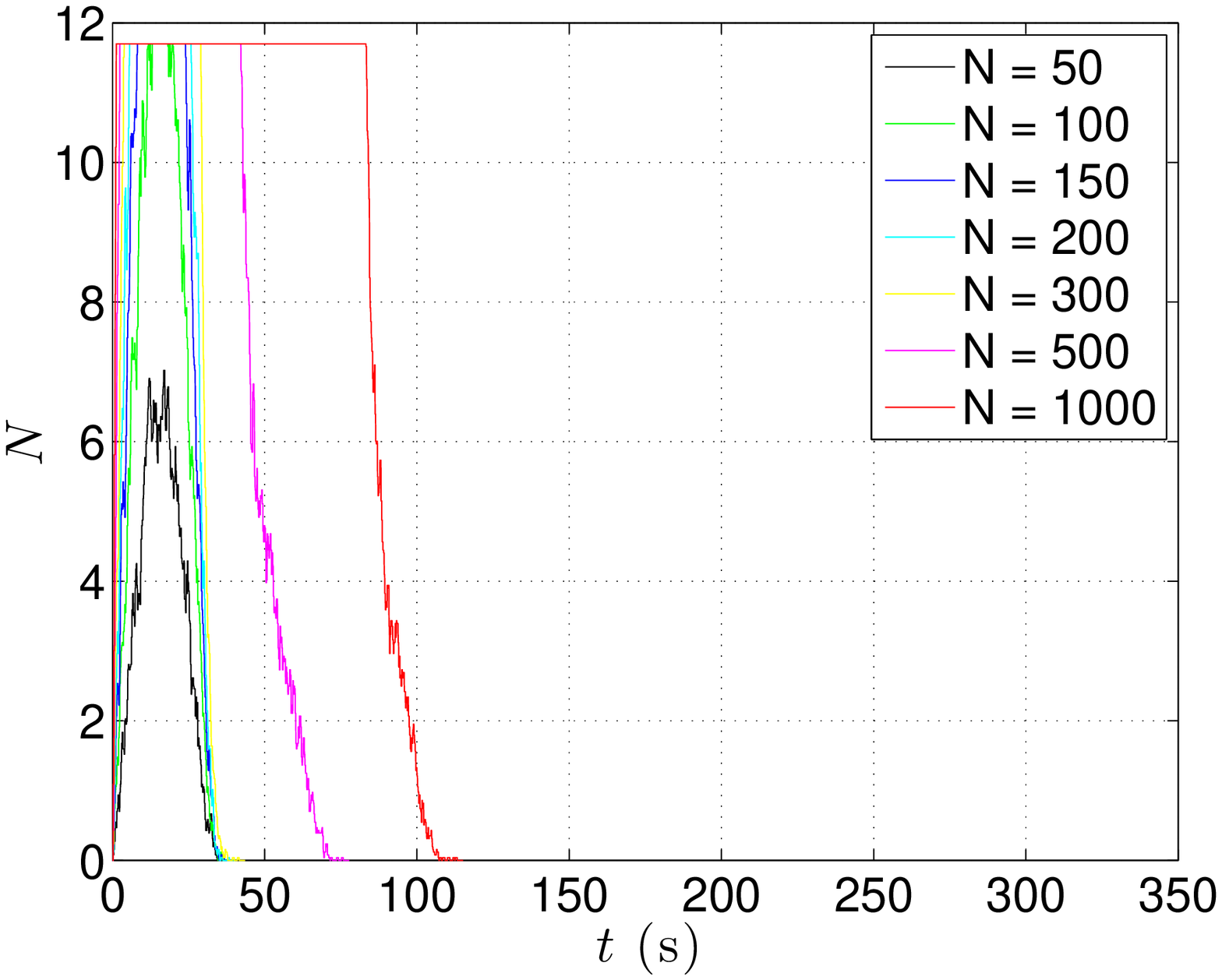}\label{fig:LatticeFlowRate}}
\subfigure[]{\includegraphics[width=6cm]{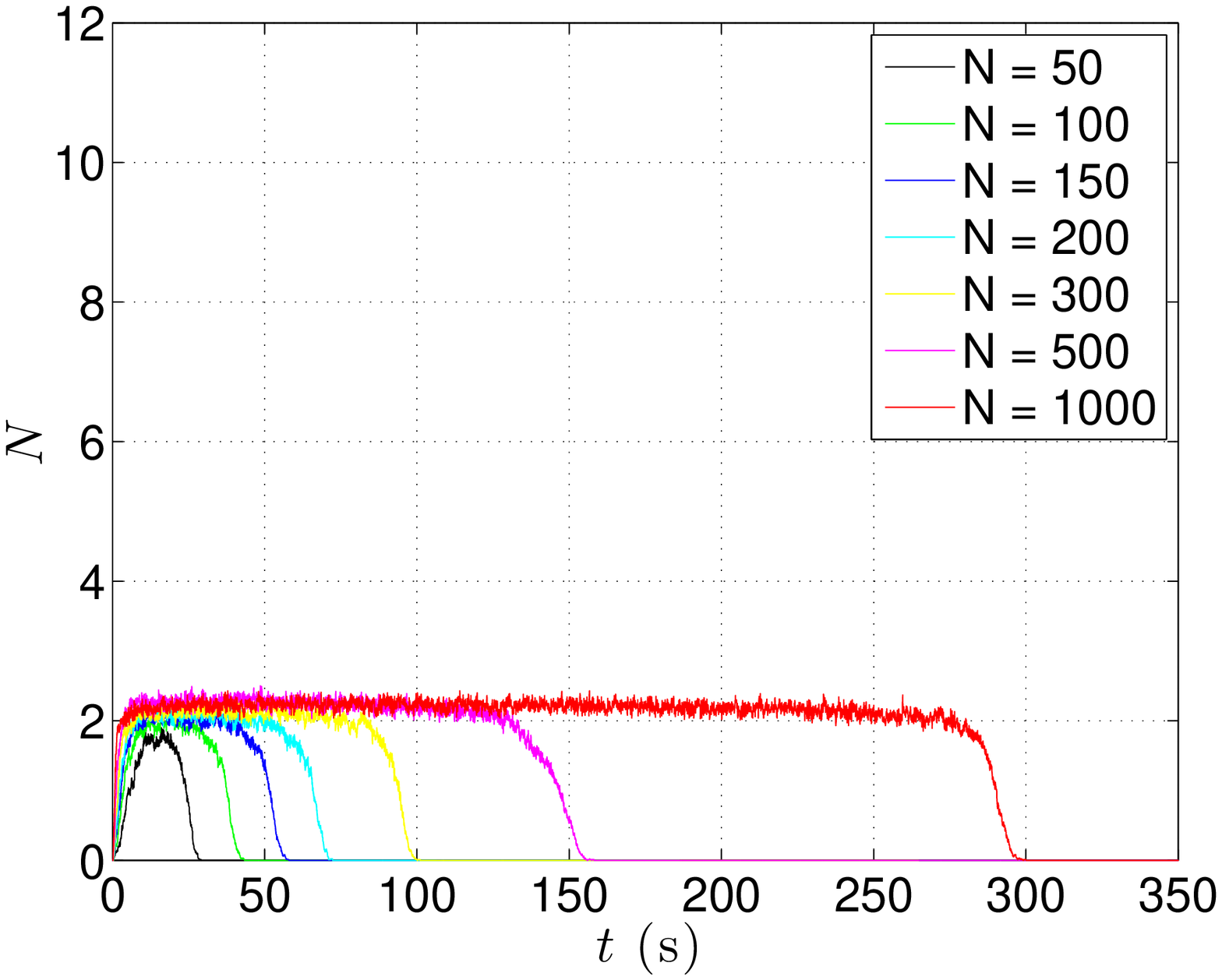}\label{fig:SocialForceFlowRate}}
\subfigure[]{\includegraphics[width=6cm]{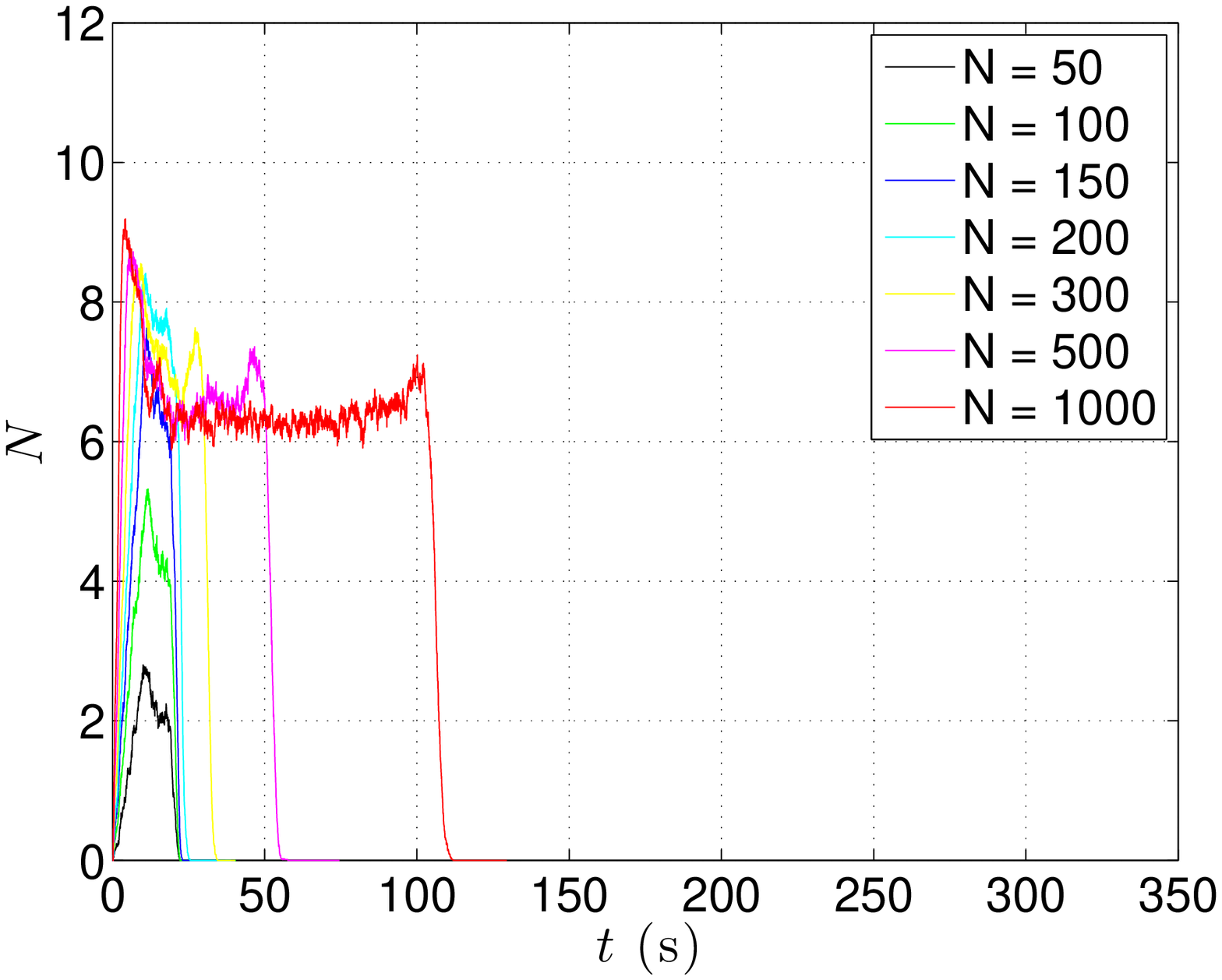}\label{fig:RVO2FlowRate}}
\caption{Flow rates at the exit as functions of time for the (a) lattice gas, (b) social force, and (c) RVO2 models. For each model and $N$, the flow rate is averaged over 100 simulations.}
\label{fig:FlowRate}
\end{figure*}

We also measured the flow rate of agents at the exit. As shown in Fig.~\ref{fig:LatticeFlowRate}, the flow rate of the lattice gas model saturates at around 11.8 persons per second due to the finite sizes of the agents and the exit. The social force model has the lowest flow rate of 2.2 people per second (Fig.~\ref{fig:SocialForceFlowRate}), because agents slow down dramatically at the end as a result of strong repulsive forces between agents. From Fig.~\ref{fig:RVO2FlowRate}), we see the same peaks before and after the congestion period for the RVO2 model, a consequence of the liquid-like dynamics of RVO2 crowds.

\begin{figure}[htbp]
\includegraphics[width=8cm]{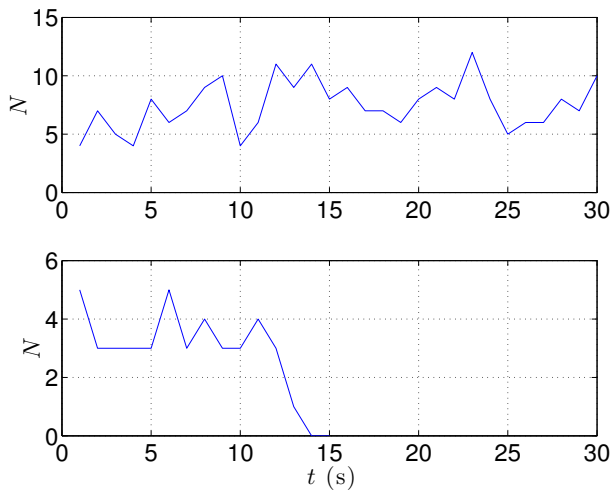}
\caption{The flow rates for evacuation during the May 2008 Sichuan Earthquake: (top) through the check-out counters of a Walmart supermarket in Chengdu, China (\url{http://www.youtube.com/watch?v=-y38ebiAnQw}); and (bottom) through the narrow exit of a classroom in an elementary school in Sichuan, China (\url{http://www.youtube.com/watch?v=e1yapP3z_L4}).}
\centering
\label{fig:RealData}
\end{figure}

In an attempt to compare simulation results with real data we obtained a number of videos showing real-life evacuations~\cite{Yang2013,Yang2011} during the Sichuan Earthquake on May 12, 2008. We analysed two such videos, \url{http://www.youtube.com/watch?v=-y38ebiAnQw} and \url{http://www.youtube.com/watch?v=e1yapP3z_L4}.
Because the geometry of the scene and the viewing angles are not known, we could only measure the flow rates on a second-by-second basis from these two videos (Fig.~\ref{fig:RealData}), for comparison against what we measured in our simulations. Unlike the computational study, where we could average over 100 simulations to get smoothly varying flow rates, the flow rates measured from the videos are noisy. However, sensible comparisons can still be made. In \url{http://www.youtube.com/watch?v=-y38ebiAnQw}, the crowd is shown to evacuate through the check-out counters of a Walmart supermarket. The flow rate was high, but there was no congestion, because of the wide exit. This flow rate cannot be compared against our simulations of evacuation through a narrow exit.

In \url{http://www.youtube.com/watch?v=e1yapP3z_L4}, students were evacuating their classroom through a narrow door. From the video, we see that the width of the door allows no more than three school children to exit simultaneously. Therefore, the exit dimensions seem to correspond to the exit dimensions use in our simulations. From Fig.~\ref{fig:RealData}, we see that the average flow rate in the congestion phase is about 3 persons per second. This is far below the congestion flow rate in the lattice gas model, and very close to the congestion flow rate of the social force model. Based on this observable, we find that the social force model agrees best with the real data. This small test also tells us that the three models can be distinguished through their congestion flow rates, if other observables could not be measured.

\subsection{DISTATIS comparison}

\begin{figure}[htbp]
\centering
\includegraphics[scale=0.5]{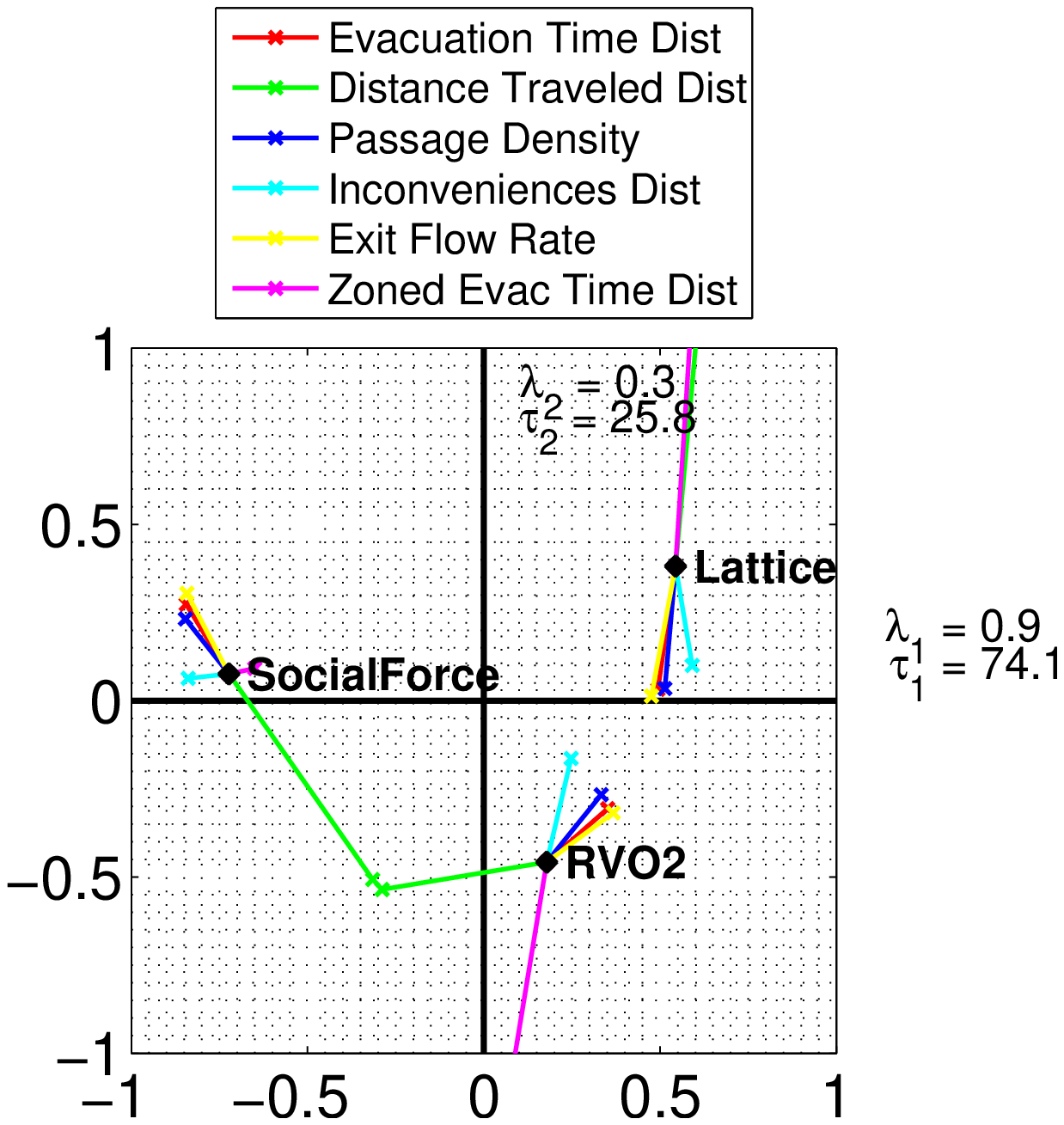}
\caption{The DISTATIS analysis of the three models with $N = 1000$ agents. In this plot, $\tau$ is the quality of compromise, and $\lambda$ is the eigenvalue.}
\label{fig:Distatis1000}
\end{figure}

Finally, we make use of the DISTATIS method explained in Sec.~\ref{sec:distatis} to compare the three models with all six different measurements. Fig.~\ref{fig:Distatis1000} shows the different measurements of the three models projected onto their first two principal components. The barycenters for the three models are also shown. From Fig.~\ref{fig:Distatis1000} we see that based on the total distance traveled (green), the social force and RVO2 models have similar distributions, which are both very different from that of the lattice gas model. On the other hand, the lattice gas and RVO2 models have similar distributions that are significantly different that of the social force model.

More importantly, we find the clustering of some observables: evacuation time, passage density, and flow rate all tells us that the lattice gas model is similar to the RVO2 model, and different from the social force model. We also find the inconvenience and zoned evacuation time provide some indication that the three models are qualitatively different. In fact, the DISTATIS method indicates that the zoned evacuation time is one key metric that clearly distinguishes the three models. Therefore, any real world experiments should attempt to measure this. This observable maximally discriminates between the three models and therefore it is reasonable to expect it will also discriminate between real-world data and the models.

\section{Conclusion}\label{Conclusions}

To conclude, we simulated emergency evacuation from a rectangular room with a single narrow exit using three models: (1) lattice gas, (2) social force, and (3) RVO2, with different starting number of agents in the room. From these simulations, we measured six observables: (1) evacuation time, (2) zoned evacuation time, (3) passage density, (4) total distance traveled, (5) inconvenience and (6) flow rate. Besides qualitative comparison of the various distributions obtained, we also compared the three models quantitatively using the DISTATIS method. Comparing the evacuation time, passage density, and flow rate distributions, we find that the lattice gas and RVO2 models are similar, both of which are very different from the social force model. On the other hand, if we compare the distribution of total distance traveled, the social force and RVO2 models are more similar to each other, and very different from the lattice gas model. We compared the simulated congestion flow rates of the three models to congestion flow rates of school children evacuating from their classroom during the May 2008 Sichuan Earthquake, and find based on this observable that the social force model agrees best with the egress data. Finally, from an analysis of the DISTATIS we have identified the zoned evacuation time as the one observable metric that can best discriminate between these models, and also between models and real-world data

\section*{Acknowledgements}

SAC thanks Zhongliang Wu and Changsheng Jiang of the Institute of Geophysics, China Earthquake Administration for discussions. CEL and SAC acknowledge support from the Nanyang Technological University New Initiatives Grant. PMAS acknowledges the support from the FET-Proactive grant TOPDRIM, number FP7-ICT-318121 and the `SimCity' eScience grant 027.012.104. MHL acknowledges funding by the Nanyang Technological University Startup Grant M58020019. PMAS also acknowledges a grant from the 'Leading Scientist Program' of the Government of the Russian Federation, under contract 11.G34.31.0019.

\bibliographystyle{IEEEtran}
\bibliography{CAMPS.bib}

\begin{thebibliography}{10}
\providecommand{\url}[1]{#1}
\csname url@samestyle\endcsname
\providecommand{\newblock}{\relax}
\providecommand{\bibinfo}[2]{#2}
\providecommand{\BIBentrySTDinterwordspacing}{\spaceskip=0pt\relax}
\providecommand{\BIBentryALTinterwordstretchfactor}{4}
\providecommand{\BIBentryALTinterwordspacing}{\spaceskip=\fontdimen2\font plus
\BIBentryALTinterwordstretchfactor\fontdimen3\font minus
  \fontdimen4\font\relax}
\providecommand{\BIBforeignlanguage}[2]{{%
\expandafter\ifx\csname l@#1\endcsname\relax
\typeout{** WARNING: IEEEtran.bst: No hyphenation pattern has been}%
\typeout{** loaded for the language `#1'. Using the pattern for}%
\typeout{** the default language instead.}%
\else
\language=\csname l@#1\endcsname
\fi
#2}}
\providecommand{\BIBdecl}{\relax}
\BIBdecl

\bibitem{Still:2000tp}
G.~K. Still, ``{Crowd Dynamics},'' Ph.D. dissertation, University of Warwick,
  University of Warwick, Department of Mathematics, Aug. 2000.

\bibitem{Zhou:2010:CMS:1842722.1842725}
\BIBentryALTinterwordspacing
S.~Zhou, D.~Chen, W.~Cai, L.~Luo, M.~Y.~H. Low, F.~Tian, V.~S.-H. Tay, D.~W.~S.
  Ong, and B.~D. Hamilton, ``Crowd modeling and simulation technologies,''
  \emph{ACM Trans. Model. Comput. Simul.}, vol.~20, no.~4, pp. 20:1--20:35,
  Nov. 2010. [Online]. Available:
  \url{http://doi.acm.org/10.1145/1842722.1842725}
\BIBentrySTDinterwordspacing

\bibitem{Gwynne1999741}
\BIBentryALTinterwordspacing
S.~Gwynne, E.~Galea, M.~Owen, P.~Lawrence, and L.~Filippidis, ``A review of the
  methodologies used in the computer simulation of evacuation from the built
  environment,'' \emph{Building and Environment}, vol.~34, no.~6, pp. 741 --
  749, 1999. [Online]. Available:
  \url{http://www.sciencedirect.com/science/article/pii/S0360132398000572}
\BIBentrySTDinterwordspacing

\bibitem{Regelous:2011vt}
S.~Regelous and K.~Mannion. (2011) {Massive Software -- Simulating Life}.

\bibitem{Reynolds:1987vm}
C.~W. Reynolds, ``{Flocks, herds and schools: A distributed behavioral
  model},'' \emph{Computer Graphics}, vol.~21, no.~4, pp. 25--34, Jul. 1987.

\bibitem{Snape:2012}
J.~Snape, ``Reciprocal collision avoidance and navigation for video games,'' in
  \emph{Game Developers Conf.}, San Francisco, Calif., March 2012.

\bibitem{ageOfEmpires:2013}
\BIBentryALTinterwordspacing
R.~E. Ensemble~Studios, Big Huge~Games, ``Age of empires,'' April 2013.
  [Online]. Available: \url{http://ageofempiresonline.com/en/}
\BIBentrySTDinterwordspacing

\bibitem{PhysRevE.51.4282}
\BIBentryALTinterwordspacing
D.~Helbing and P.~Moln\'ar, ``Social force model for pedestrian dynamics,''
  \emph{Phys. Rev. E}, vol.~51, pp. 4282--4286, May 1995. [Online]. Available:
  \url{http://link.aps.org/doi/10.1103/PhysRevE.51.4282}
\BIBentrySTDinterwordspacing

\bibitem{Viswanathan:ut}
V.~Viswanathan and M.~Lees, ``{Modeling and Analyzing the Human Cognitive
  Limits for Perception in Crowd Simulation},'' in \emph{Transactions on
  Computational Science}, M.~L. Gavrilova, K.~C. Tan, and C.-V. Phan,
  Eds.\hskip 1em plus 0.5em minus 0.4em\relax Springer, 2012, pp. 1--20.

\bibitem{Guy:2010uv}
\BIBentryALTinterwordspacing
S.~J. Guy, J.~Chhugani, S.~Curtis, P.~Dubey, M.~Lin, and D.~Manocha,
  ``Pledestrians: a least-effort approach to crowd simulation,'' in
  \emph{Proceedings of the 2010 ACM SIGGRAPH/Eurographics Symposium on Computer
  Animation}, ser. SCA '10.\hskip 1em plus 0.5em minus 0.4em\relax
  Aire-la-Ville, Switzerland, Switzerland: Eurographics Association, 2010, pp.
  119--128. [Online]. Available:
  \url{http://dl.acm.org/citation.cfm?id=1921427.1921446}
\BIBentrySTDinterwordspacing

\bibitem{Klupfel:2005to}
H.~Kl\"{u}pfel, M.~Schreckenberg, and T.~Meyer-K\"{o}nig,
  ``\BIBforeignlanguage{English}{Models for crowd movement and egress
  simulation},'' in \emph{\BIBforeignlanguage{English}{Traffic and Granular
  Flow '03}}, S.~Hoogendoorn, S.~Luding, P.~Bovy, M.~Schreckenberg, and
  D.~Wolf, Eds.\hskip 1em plus 0.5em minus 0.4em\relax Springer Berlin
  Heidelberg, 2005, pp. 357--372.

\bibitem{PEDFull:2011}
\BIBentryALTinterwordspacing
J.~D.~A. Richard D.~Peacock, Erica D.~Kuligowski, \emph{Pedestrian And
  Evacuation Dynamics}.\hskip 1em plus 0.5em minus 0.4em\relax Springer, 2011.
  [Online]. Available:
  \url{http://link.springer.com/book/10.1007/978-1-4419-9725-8/page/1}
\BIBentrySTDinterwordspacing

\bibitem{Mordvintsev:2012}
A.~Mordvintsev, V.~Krzhizhanovskaya, M.~Lees, and P.~M.~A. Sloot, ``Simulation
  of city evacuation coupled to flood dynamics,'' {P}edestrian and Evacuation
  Dynamics 2012, in press. Available at
  \url{http://www.science.uva.nl/research/pscs/papers/archive/Mordvintsev2012a.pdf}.

\bibitem{Henderson:1974ve}
L.~Henderson, ``{On the fluid mechanics of human crowd motion},''
  \emph{Transportation research}, vol.~8, pp. 509--515, 1974.

\bibitem{WattsJr:1987tx}
J.~M. Watts~Jr, ``{Computer models for evacuation analysis},'' \emph{Fire
  Safety Journal}, vol.~12, pp. 237--245, 1987.

\bibitem{CGF:CGF1090}
\BIBentryALTinterwordspacing
S.~Paris, J.~Pettré, and S.~Donikian, ``Pedestrian reactive navigation for
  crowd simulation: a predictive approach,'' \emph{Computer Graphics Forum},
  vol.~26, no.~3, pp. 665--674, 2007. [Online]. Available:
  \url{http://dx.doi.org/10.1111/j.1467-8659.2007.01090.x}
\BIBentrySTDinterwordspacing

\bibitem{HuNan2013}
S.~Z. Nan~Hu, Michael~Lees, Ed., \emph{A Pattern-based Modeling Framework for
  Simulating Human-like Pedestrian Steering Behaviors}.\hskip 1em plus 0.5em
  minus 0.4em\relax ACM, 2013 (to appear).

\bibitem{JOHANSSON2008}
\BIBentryALTinterwordspacing
A.~Johansson, D.~Helbing, H.~Z. Al-abideen, and S.~Al-bosta, ``From crowd
  dynamics to crowd safety: A video-based analysis,'' \emph{Advances in Complex
  Systems}, vol.~11, no.~04, pp. 497--527, 2008. [Online]. Available:
  \url{http://www.worldscientific.com/doi/abs/10.1142/S0219525908001854}
\BIBentrySTDinterwordspacing

\bibitem{Okazaki:1993wh}
S.~Okazaki and S.~Matsushita, ``{A study of simulation model for pedestrian
  movement with evacuation and queuing},'' \emph{Engineering for Crowd Safety},
  1993.

\bibitem{Ondrej:2010hv}
J.~Ond{\v r}ej, J.~Pettr{\'e}, A.-H. Olivier, and S.~Donikien, ``{A
  Synthetic-Vision Based Steering Approach for Crowd Simulation},'' \emph{ACM
  Transactions on Graphics (TOG) - Proceedings of ACM SIGGRAPH 2010}, vol.~29,
  no.~4, Jul. 2010.

\bibitem{vandenBerg:2011ww}
J.~van~den Berg, S.~J. Guy, and M.~C. Lin, ``{Reciprocal n-Body Collision
  Avoidance},'' \emph{Robotics Research}, 2011.

\bibitem{Guy:2010ko}
S.~J. Guy, J.~van~den Berg, and M.~C. Lin, ``{Geometric methods for multi-agent
  collision avoidance},'' in \emph{Symposium on Computer Graphics}, University
  of North Carolina.\hskip 1em plus 0.5em minus 0.4em\relax Utah, USA:
  Proceedings of the 2010 annual symposium on Computational geometry, Jun.
  2010, pp. 115--116.

\bibitem{vandenBerg:2008fu}
J.~van~den Berg, M.~C. Lin, and D.~Manocha, ``{Reciprocal Velocity Obstacles
  for real-time multi-agent navigation},'' in \emph{Robotics and Automation,
  2008. ICRA 2008. IEEE International Conference on}, 2008, pp. 1928--1935.

\bibitem{Hughes:2003}
R.~Hughes, ``\BIBforeignlanguage{{English}}{{The flow of human crowds}},''
  \emph{\BIBforeignlanguage{{English}}{{Annual Review Of Fluid Mechanics}}},
  vol.~{35}, pp. {169--182}, {2003}.

\bibitem{Helbing:2012}
\BIBentryALTinterwordspacing
D.~Helbing and P.~Mukerji, ``\BIBforeignlanguage{English}{Crowd disasters as
  systemic failures: analysis of the love parade disaster},''
  \emph{\BIBforeignlanguage{English}{EPJ Data Science}}, vol.~1, no.~1, pp.
  1--40, 2012. [Online]. Available: \url{http://dx.doi.org/10.1140/epjds7}
\BIBentrySTDinterwordspacing

\bibitem{Hoekstra:2010}
A.~G. Hoekstra, J.~Kroc, and P.~M.~A. Sloot, Eds., \emph{Simulating Complex
  Systems by Cellular Automata}, ser. Understanding Complex Systems.\hskip 1em
  plus 0.5em minus 0.4em\relax Springer, 2010, iSBN: 978-3-642-12202-6.

\bibitem{Marconi:2002ue}
S.~Marconi, ``{Mesoscopical modelling of complex systems},'' Ph.D.
  dissertation, Universit{\'e} de Gen{\`e}ve, Gen{\`e}ve, 2002.

\bibitem{Marconi2002}
\BIBentryALTinterwordspacing
S.~Marconi and B.~Chopard, ``\BIBforeignlanguage{English}{A multiparticle
  lattice gas automata model for a crowd},'' in
  \emph{\BIBforeignlanguage{English}{Cellular Automata}}, ser. Lecture Notes in
  Computer Science, S.~Bandini, B.~Chopard, and M.~Tomassini, Eds.\hskip 1em
  plus 0.5em minus 0.4em\relax Springer Berlin Heidelberg, 2002, vol. 2493, pp.
  231--238. [Online]. Available:
  \url{http://dx.doi.org/10.1007/3-540-45830-1_22}
\BIBentrySTDinterwordspacing

\bibitem{Nagai:2004kl}
R.~Nagai, T.~Nagatani, M.~Isobe, and T.~Adachi, ``{Effect of Exit Configuration
  on Evacuation of a Room Without Visibility},'' \emph{Physica A: Statistical
  Mechanics and its Applications}, vol. 343, pp. 712--724, Nov. 2004.

\bibitem{nishinari2004extended}
K.~Nishinari, A.~Kirchner, A.~Namazi, and A.~Schadschneider, ``Extended floor
  field ca model for evacuation dynamics,'' \emph{IEICE Transactions on
  information and systems}, vol.~87, no.~3, pp. 726--732, 2004.

\bibitem{Henein:2006jq}
C.~M. Henein and T.~White, \emph{{Crowds and Cellular Automata}}, ser. Lecture
  Notes in Computer Science, D.~Hutchison, T.~Kanade, J.~Kittler, J.~M.
  Kleinberg, F.~Mattern, J.~C. Mitchell, M.~Naor, O.~Nierstrasz,
  C.~Pandu~Rangan, B.~Steffen, M.~Sudan, D.~Terzopoulos, D.~Tygar, M.~Y. Vardi,
  G.~Weikum, S.~Yacoubi, B.~Chopard, and S.~Bandini, Eds.\hskip 1em plus 0.5em
  minus 0.4em\relax Berlin, Heidelberg: Springer Berlin Heidelberg, 2006, vol.
  4173.

\bibitem{Tajima:2001to}
Y.~Tajima and T.~Nagatani, ``{Scaling behavior of crowd flow outside a hall},''
  \emph{Physica A}, vol. 292, pp. 545--554, 2001.

\bibitem{isobe2004experiment}
M.~Isobe, T.~Adachi, and T.~Nagatani, ``Experiment and simulation of pedestrian
  counter flow,'' \emph{Physica A: Statistical Mechanics and its Applications},
  vol. 336, no.~3, pp. 638--650, 2004.

\bibitem{Kamphuis:2004uu}
A.~Kamphuis and M.~H. Overmars, ``{Finding paths for coherent groups using
  clearance},'' in \emph{Proceedings of the 2004 ACM SIGGRAPH Symposium on
  Computer Animation}.\hskip 1em plus 0.5em minus 0.4em\relax Utrecht
  University, 2004.

\bibitem{Xi:2010uc}
H.~Xi, S.~Lee, and Y.-J. Son, ``{An Integrated Pedestrian Behavior Model Based
  on Extended Decision Field Theory and Social Force Model},'' in \emph{2010
  Winter Simulation Conference}, Nov. 2010, pp. 824--836.

\bibitem{Peng:2009cc}
G.~Peng and X.~Ruihua, ``{3-Tier Architecture for Pedestrian Agent in Crowd
  Simulation},'' \emph{Pedestrian and Evacuation Dynamics 2008}, no. Chapter
  53, pp. 585--595, Dec. 2009.

\bibitem{Helbing:2000ef}
D.~Helbing, I.~Farkas, and T.~Vicsek, ``{Simulating Dynamical Features of
  Escape Panic},'' \emph{Physical Review E}, vol. cond-mat.stat-mech, Sep.
  2000.

\bibitem{Fiorini:1993hi}
P.~Fiorini and Z.~Shiller, ``{Motion planning in dynamic environments using the
  relative velocity paradigm},'' in \emph{Robotics and Automation, 1993.
  Proceedings., 1993 IEEE International Conference on}, 1993, pp. 560--565.

\bibitem{vandenBerg:2008cq}
J.~van~den Berg, S.~Patil, J.~Sewall, D.~Manocha, and M.~C. Lin, ``{Interactive
  Navigation of Multiple Agents in Crowded Environments},'' in \emph{2008
  symposium on Interactive 3D graphics and games}.\hskip 1em plus 0.5em minus
  0.4em\relax University of North Carolina, 2008.

\bibitem{Guy:2009gu}
S.~J. Guy, J.~Chhugani, C.~Kim, N.~Satish, M.~C. Lin, D.~Manocha, and P.~Dubey,
  ``{ClearPath: highly parallel collision avoidance for multi-agent
  simulation},'' in \emph{SCA '09: Proceedings of the 2009 ACM
  SIGGRAPH/Eurographics Symposium on Computer Animation}.\hskip 1em plus 0.5em
  minus 0.4em\relax ACM Request Permissions, Aug. 2009.

\bibitem{Guy:2010te}
S.~J. Guy, M.~C. Lin, and D.~Manocha, ``{Modelling Collision Avoidance Behavior
  for Virtual Humans},'' in \emph{9th Int. Conf. on Autonomous Agents and
  Multiagent Systems (AAMAS 2010)}, Toronto, Canada, May 2010, pp. 575--582.

\bibitem{Luke:2005wc}
S.~Luke, C.~Cioffi-Revilla, L.~Panait, K.~Sullivan, and G.~Balan, ``{MASON: A
  Multi-Agent Simulation Environment},'' \emph{Simulation : Transactions of the
  society for Modeling and Simulation International}, vol.~82, no.~7, pp.
  517--527, 2005.

\bibitem{Pan:2006vp}
X.~Pan, ``{Computational Modeling of Human and Social Behaviors For Emergency
  Egress Analysis},'' Ph.D. dissertation, Stanford University, The Department
  of Civil and Environmental Engineering, Stanford University, Jun. 2006.

\bibitem{fruin1992designing}
J.~Fruin, ``Designing for pedestrians,'' \emph{Public Transportation United
  States}, 1992.

\bibitem{Bauer2011}
\BIBentryALTinterwordspacing
D.~Bauer, ``\BIBforeignlanguage{English}{Comparing pedestrian movement
  simulation models for a crossing area based on real world data},'' in
  \emph{\BIBforeignlanguage{English}{Pedestrian and Evacuation Dynamics}},
  R.~D. Peacock, E.~D. Kuligowski, and J.~D. Averill, Eds.\hskip 1em plus 0.5em
  minus 0.4em\relax Springer US, 2011, pp. 547--556. [Online]. Available:
  \url{http://dx.doi.org/10.1007/978-1-4419-9725-8_49}
\BIBentrySTDinterwordspacing

\bibitem{Seyfried2008}
\BIBentryALTinterwordspacing
A.~Seyfried and A.~Schadschneider, ``Fundamental diagram and validation of
  crowd models,'' in \emph{Cellular Automata}, ser. Lecture Notes in Computer
  Science, H.~Umeo, S.~Morishita, K.~Nishinari, T.~Komatsuzaki, and S.~Bandini,
  Eds.\hskip 1em plus 0.5em minus 0.4em\relax Springer Berlin Heidelberg, 2008,
  vol. 5191, pp. 563--566. [Online]. Available:
  \url{http://dx.doi.org/10.1007/978-3-540-79992-4_77}
\BIBentrySTDinterwordspacing

\bibitem{Lin:1991it}
J.~Lin, ``{Divergence measures based on the Shannon entropy},'' \emph{IEEE
  Transactions on Information Theory}, vol.~37, no.~1, pp. 145--151, 1991.

\bibitem{Abdi:DISTATIS}
H.~Abdi, D.~Valentin, U.~D. Bourgogne, and B.~Edelman, ``Distatis: The analysis
  of multiple distance matrices,'' in \emph{Proceedings of the IEEE Computer
  Society: International Conference on Computer Vision and Pattern
  Recognition}, 2005, pp. 42--47.

\bibitem{Yang2013}
\BIBentryALTinterwordspacing
X.~Yang and Z.~Wu, ``\BIBforeignlanguage{English}{Civilian monitoring video
  records for earthquake intensity: a potentially unbiased online information
  source of macro-seismology},'' \emph{\BIBforeignlanguage{English}{Natural
  Hazards}}, vol.~65, no.~3, pp. 1765--1781, 2013. [Online]. Available:
  \url{http://dx.doi.org/10.1007/s11069-012-0447-3}
\BIBentrySTDinterwordspacing

\bibitem{Yang2011}
\BIBentryALTinterwordspacing
X.~Yang, Z.~Wu, and Y.~Li, ``Difference between real-life escape panic and
  mimic exercises in simulated situation with implications to the statistical
  physics models of emergency evacuation: The 2008 wenchuan earthquake,''
  \emph{Physica A: Statistical Mechanics and its Applications}, vol. 390,
  no.~12, pp. 2375 -- 2380, 2011. [Online]. Available:
  \url{http://www.sciencedirect.com/science/article/pii/S0378437110008770}
\BIBentrySTDinterwordspacing

\end{thebibliography}

\end{document}